\documentclass[12pt,psamsfonts,reqno]{amsart}

\theoremstyle{definition}

\theoremstyle{remark}

\newtheorem*{remark*}{Remark}

\newcommand{\vev}[1]{\left\langle #1 \right\rangle}
\newcommand{\ket}[1]{\left |  #1 \right \rangle}
\newcommand{\bra}[1]{\left \langle  #1 \right |}

\newcommand {\CalO} {\mathcal O}

\newcommand {\CalN} {\mathcal N}

\newcommand {\CalY} {\mathcal Y}
\newcommand {\CalX} {\mathcal X}

\newcommand {\BZ}   {\mathbb Z}
\newcommand {\BC}   {\mathbb C}

\newcommand {\BP}   {\mathbb P}

\newcommand{\bc}{\mathbf{c}}
\newcommand{\bn}{\mathbf{n}}

\newcommand{\bI}{\mathbf{I}}
\newcommand{\bv}{\mathbf{v}}

\newcommand{\bW}{\mathbf{W}}
\newcommand{\bw}{\mathbf{w}}

\newcommand{\bH}{\mathbf{H}}
\newcommand{\bM}{\mathbf{M}}
\newcommand{\bN}{\mathbf{N}}
\newcommand{\bK}{\mathbf{K}}
\newcommand{\bQ}{\mathbf{Q}}

\newcommand{\bX}{\mathbf{X}}

\newcommand{\bY}{\mathbf{Y}}

\newcommand{\msS}{\mathscr{S}}
\newcommand{\si}{\mathsf{i}}
\newcommand{\sn}{\mathsf{n}}
\newcommand{\sw}{\mathsf{w}}

\newcommand{\sm}{\mathsf{m}}
\newcommand{\sk}{\mathsf{k}}
\newcommand{\st}{\mathsf{t}}

\newcommand{\sS}{\mathsf{S}}
\newcommand{\sV}{\mathsf{V}}
\newcommand{\sT}{\mathsf{T}}

\newcommand{\sY}{\mathsf{Y}}

\newcommand{\tT}{T}

\newcommand{\g}{\mathfrak{g}}

\newcommand{\frakM}{\mathfrak{M}}
\newcommand{\fc}{\mathfrak{c}}
\newcommand{\fq}{\mathfrak{q}}
\newcommand {\p} {\partial}

\newcommand{\ep}{\epsilon}

\DeclareMathOperator{\ch}{ch}

\DeclareMathOperator{\Rep}{Rep}
\DeclareMathOperator{\Hom}{Hom}
\DeclareMathOperator{\Ext}{Ext}

\DeclareMathOperator{\Tr} {Tr}

\DeclareMathOperator{\rk} {rk}

\numberwithin{equation}{section}

\newcommand{\spacetime}{\mathcal{S}}

\usepackage[margin=1in]{geometry}

\usepackage{amsmath}
\usepackage{amssymb}
\usepackage{amsxtra}
\usepackage{amscd}
\usepackage[mathscr]{euscript}

\usepackage[pagebackref=false]{hyperref}
\hypersetup{colorlinks = false, ,extension = notused,linkcolor = blue, anchorcolor = red,citecolor = blue,filecolor = red,pagecolor = red,urlcolor = blue}
\usepackage[numbers,sort&compress]{natbib}
\usepackage{hypernat}

\usepackage{iftex}
\ifxetex
        \usepackage{fontspec}
\else
\fi

\usepackage{float}
\usepackage{tikz-cd}
\usepackage[mathscr]{euscript}

\begin{document}

\title{Quiver W-algebras}

\author{Taro Kimura}
\author{Vasily Pestun}

\address{Taro Kimura, Keio University, Japan}
\address{Vasily Pestun, IHES, France} 

\begin{abstract} 
 For a quiver with weighted arrows we define gauge-theory K-theoretic W-algebra generalizing the definition of Shiraishi et al., and Frenkel and Reshetikhin.
 In particular, we show that the $qq$-character construction of gauge theory presented by Nekrasov is isomorphic to the definition of the W-algebra in the operator formalism as a commutant of screening charges in the free field representation.
 Besides, we allow arbitrary quiver and expect interesting applications to representation theory of generalized Borcherds--Kac--Moody Lie algebras, their quantum affinizations and associated W-algebras. 
\end{abstract}

\maketitle 

\tableofcontents

\parskip=4pt

\section{Introduction}\label{se:Introduction}

 Let $\Gamma$ be a quiver with $\mu$-weighted arrows $\mu: \Gamma_1 \to \BC^{\times}$.
 We construct two-parametric algebra $W_{q_1,q_2}(\Gamma)$ from the equivariant K-theory on the moduli space of $\Gamma$-quiver sheaves on $\BC^2_{q_1, q_2}$.
 If a quiver is simply-laced Dynkin graph, our construction agrees with Frenkel--Reshetikhin's definition of $W_{q_1,q_2}(\g_{\Gamma})$~\cite{Frenkel:1997} on the one hand.
 On the other hand, it explains that certain observables of the gauge theory, coming from the $q_2$-lift of the gauge-theory construction of $q_1$-characters in \cite{Nekrasov:2013xda}, described in details and called $q_1 q_2$-characters in \cite{Nekrasov:2015wsu}, can be promoted to \emph{operator valued currents} that form non-commutative associative 
current algebra $W_{q_1,q_2}(\g_{\Gamma})$. 

 In our constructions the $q_1q_2$-character currents are valued in the algebra of differential operators in \emph{higher times} $t_{i,1}, \dots, t_{i,\infty}$ of the gauge theory \cite{Marshakov:2006ii}. 
 We show that the algebra of these differential operators is equivalent to the Heisenberg algebra of $q_1 q_2$-bosons used by Shiraishi et al.~\cite{Shiraishi:1995rp} and Frenkel--Reshetikhin~\cite{Frenkel:1996,Frenkel:1997} to define $W_{q_1,q_2}(\g_{\Gamma})$ algebra.
 Specializing to $\g_{\Gamma}$ of ADE type (in this case $\Gamma$ has no loops and hence $\mu$-parameters represent necessarily the trivial class in $H_1(\Gamma, \BC^{\times})$ and hence are gauged away)  we  show that the pole cancellation construction of \cite{Nekrasov:2013xda} and \cite{Nekrasov:2015wsu} developed from the cut cancellation construction of \cite{Nekrasov:2012xe}, is isomorphic to the definition of $W_{q_1, q_2}(\g_{\Gamma})$ algebra in \cite{Frenkel:1997} as the commutant of \emph{screening charges}, hence explaining the isomorphism between the gauge theory construction presented in \cite{Nekrasov:2015wsu} and the algebraic contruction of \cite{Frenkel:1997}.

 The gauge theory definition of $W_{q_1, q_2}(\Gamma)$ algebra is symmetric in exchange $q_1 \leftrightarrow q_2$.
 However, for the free field realization there is a choice between $(q_1,q_2)$ and $(q_2,q_1)$.
 The equivalence between the two realizations, transparent from the gauge theory, leads to `quantum $q$-geometric' Langlands equivalence.
 The duality of W-algebras and the connection with geometric Langlands was first found in \cite{Feigin:1992affine}. 
 The `quantum $q$-geometric' Langlands \cite{Frenkel:1997, Frenkel:2010wd} degenerates to the (CFT) `quantum geometric' Langlands duality $ \beta \leftrightarrow \beta^{-1}$ in the limit $q_1=e^{\ep_1}, q_2=e^{\ep_2}$ with $\ep_1, \ep_2 \to 0$ and $\beta = -\ep_1/\ep_2$, and further down in the limit $\ep_1 = \hbar, \ep_2 = 0$ to the `geometric' Langlands duality \cite{Beilinson:Drinfeld, Kapustin:2006pk, Nekrasov:2010ka}. 
 For a survey of duality of W-algebras and its connection with the geometric Langlands program see \cite{Frenkel:1997} and
\cite{frenkel2007lectures} section 8.6. 

 In the language of complex integrable systems, and in the reverse order, the `geometric' Langlands duality $(\ep_1 = \hbar, \ep_2=0)$ is T-duality along the fibers of the phase space of Hitchin integrable system.
 The K-theory lift to $q_1=e^{\hbar}, q_2 = 1$ converts Hitchin integrable system on flat curve $C$ to group valued Hitchin integrable system on $C$ \cite{Hurtubise:2002}, equivalently to the integrable system of periodic monopoles on $C \times S^1$ \cite{Nekrasov:2012xe}. 
 In the limit $q_2 = 1$ there is isomorphism, found in \cite{Frenkel:1998}, between the algebra $W_{q_1,q_2}(\g_{\Gamma})$, which turns into commutative algebra, and the K-theory ring of the category of certain representations of the quantum loop group $\mathbf{U}_{q_1}(L \g_{\Gamma})$
 \begin{equation}
W_{q_1, 1}(\g_{\Gamma}) \simeq K(\Rep  \mathbf{U}_{q_1}(L \g_\Gamma))
 \end{equation}
 The character of the R-matrix maps the elements of $W_{q_1, 1}(\g_{\Gamma})$ to the commuting Hamiltonians of quantum integrable system \cite{Nekrasov:2013xda, Frenkel:1998, Frenkel:2013uda}. 

 The geometric realization of $\mathbf{U}_{q_1}(L\g_{\Gamma})$ and of $q_1$-characters in $W_{q_1, 1}(\g_{\Gamma})$ was obtained by Nakajima~\cite{Nakajima:1999} after Ringel~\cite{Ringel}, Lusztig~\cite{Lusztig}, Ginzburg--Vasserot~\cite{MR1208827} from the equivariant K-theory of the $\BC^{\times}_{q_1}$-equivariant cotangent bundle of the moduli space $T^{*}_{q_1} \frakM(\Gamma, \BC\mathbf{Mod})$ of $\Gamma$-quiver representations in the category of vector spaces $\BC$-$\mathbf{Mod}$.
 To see the $q_2$-parameter one needs to consider the central extension of $\mathbf{U}_{q_1}(L\g_{\Gamma})$ to $\mathbf{U}_{q_1}(\widehat \g_{\Gamma})$ (quantum Drinfeld affinization of $\g_{\Gamma}$).
 The central extension is missing in Nakajima's construction which concerns only the specialization of $\mathbf{U}_{q_1}(\widehat \g_{\Gamma})$ by the trivial center to $\mathbf{U}_{q_1}(L\g_{\Gamma})$.%
\footnote{The parameters $(q_1^{\frac 1 2},q_2^{-\frac 1 2})$ are called by $(q, t)$ in~\cite{Shiraishi:1995rp, Frenkel:1997, Frenkel:2010wd}.
 However, the parameter $t$ in Nakajima's $(q,t)$-characters \cite{Nakajima:2004}, which grades the cohomological degree, has different meaning from the present $q_2$.}

 Compared to Nakajima, we replace a point by a complex variety $\spacetime$ and replace $\BC\mathbf{Mod}$ by $\CalO_{\spacetime}\mathbf{Mod}$, so that we consider equivariant K-theory on the moduli space $\frakM(\Gamma,\CalO_{\spacetime}\mathbf{Mod})$ of $\Gamma$-quiver representations in the category of coherent sheaves on a complex variety $\spacetime$.
 For $\spacetime = \BC^2_{q_1, q_2}$ we recover $W_{q_1,q_2}(\g_{\Gamma})$ from K-theory on $\frakM(\Gamma, \CalO_{\BC_{q_1,q_2}}\mathbf{Mod})$.
 As proposed by Nekrasov in~\cite{Nekrasov:2015wsu} using complex 4-dimensional setup this should be  equivalent to considering the K-theory on the $q_2$-twisted fiber-parity-inversed total space of the tangent bundle to Nakajima's quiver variety $\Pi T_{q_2} T^{*}_{q_1 q_2} \frakM(\Gamma,\BC\mathbf{Mod})$.

 We expect that K-theory definition of quiver W-algebra $W(\Gamma, \BC^2_{q_1, q_2})$ can be given in a more geometric sense in the more general situation when $\BC^2_{q_1, q_2}$ is replaced by a generic complex variety $\spacetime$ factorized into a product $\spacetime = \spacetime_1 \times \spacetime_2$ and we expect that the duality $\spacetime_1 \leftrightarrow \spacetime_2$ will be lifted to higher Langlands duality.
 The relation to cohomological Hall algebra of quiver~\cite{Kontsevich:2010px} remains to be clarified. 

 This paper takes equivariant K-theory as example of  generalized cohomological theory corresponding to the supersymmetric 5d theory reduced on $S^1$.
 However, all constructions remain intact if equivariant K-theory is replaced by the ordinary equivariant cohomology (4d theory) or by equivariant elliptic cohomology (6d theory reduced on elliptic curve).
 Consequently, the geometric construction of K-theoretic W-algebra can be scaled to its Yangian version \cite{Frenkel:1997, Frenkel:1998, Frenkel:2013uda, Hou:1997yt} using cohomology and lifted to the elliptic version \cite{Iqbal:2015fvd,Nieri:2015dts} using elliptic cohomology.

 For $A_{r}$-quivers the defining relation of the present note between gauge theory and $W(A_{r})$-algebra after the $90^\circ$ brane rotation (the exchange between the rank of the gauge group in the quiver nodes and the rank of the quiver \cite{Witten:1997sc}, equivalently Nahm transform, fiber-base duality, S-duality) implies the AGT duality of~\cite{Alday:2009aq,Wyllard:2009hg,Braverman:2014xca}.
 The invariance under the brane rotation of the gauge theory partition function is clear from the formalism of refined topological vertex \cite{Iqbal:2007ii, Awata:2008ed,Nekrasov:2014nea} and was explicitly checked in \cite{Bao:2011rc}.
 The relation between quiver gauge theories and W-algebras in terms of Toda conformal blocks for finite ADE quivers also appeared in \cite{Aganagic:2015cta}. 
 Also it would be interesting to interpret the higher times and the meaning of the presented $W(\Gamma)$-symmetry in the context of topological string on toric Calabi--Yau realization of the gauge theory partition function for ADE and affine ADE quivers \cite{Katz:1997eq}.

 We do not restrict $\Gamma$ to be quiver of finite Dynkin type and consequently expect interesting applications to representation theory of (quantum affinization of) generalized Borcherds--Kac--Moody Lie algebras, such as $E_{11}$ symmetry prominently appearing in M-theory or Borcherds Monster Lie algebra for Conway--Norton moonshine.
 The affine and hyperbolic quivers generate new W-algebras describing affine (such as sinh-Gordon) and hyperbolic quantum 2d Toda models.

\subsubsection*{Notes}
 The origins of the $q$-Virasoro symmetry the case of single node quiver $\Gamma = A_1 = \circ$ to $\mathrm{Vir}_{q_1,q_2}= W_{q_1,q_2}(A_1)$ algebra can be traced to Eynard's $q$-deformed single matrix  model~\cite{Eynard:2008mt,Sulkowski:2009ne,Nedelin:2015mio}.
 Elliptic version of matrix model is discussed in \cite{Mironov:2015thk}. 
 It would be interesting to explore $\Gamma$-quiver matrix models beyond Dynkin graphs of finite and affine type~\cite{Kostov:1992ie}.

 The regularity of $qq$-characters was explained by Nekrasov in the talk in \href{http://physics.princeton.edu/strings2014/Talk_titles.shtml}{Strings 2014} and multiple other talks.
 In \cite{Bourgine:2015szm} the regularity for a linear quiver was interpreted in the language of the quantum toroidal algebra $\mathbf{U}_{q_1,q_2}(\widehat{\widehat{\mathfrak{gl}}}_1)$ which by Nakajima's construction~\cite{Nakajima1994, Ginzburg:1995b,Kapranov-Vasserot_2000,Schiffmann:2009,Varagnolo:1999,Schiffmann:2012b,Maulik:2012} acts on the instanton moduli spaces on $\BC^2_{q_1,q_2}$ for each individual node.

 \subsection*{Acknowledgements}
 
 We thank for discussions and comments Alexey Sevastyanov, Edward Frenkel, Nikita Nekrasov and Samson Sha\-ta\-shvili.
 TK is grateful to Institut des Hautes \'Etudes Scientifiques for hospitality where a part of this work has been done.
The work of TK was supported in part by Keio Gijuku Academic Development Funds, JSPS Grant-in-Aid for Scientific Research (Nos.~JP13J04302 and JP17K18090), the MEXT-Supported Program for the Strategic Research Foundation at Private Universities ``Topological Science'' (No.~S1511006), JSPS Grant-in-Aid for Scientific Research on Innovative Areas ``Topological Materials Science'' (No.~JP15H05855), and ``Discrete Geometric Analysis for Materials Design'' (No.~JP17H06462).
VP acknowledges grant RFBR 15-01-04217 and RFBR 16-02-01021. The research of VP on this project has received funding from the European Research Council (ERC) under the European Union's Horizon 2020 research and innovation program (QUASIFT grant agreement 677368).

\section{Quiver gauge theory}

In this section we define the extended partition function of quiver gauge theories \cite{Douglas:1996sw,Nekrasov:2002qd,Marshakov:2006ii}. 



\subsection{Quiver}

 Let $\Gamma$ be a quiver with the set of nodes $\Gamma_0$ and the set of arrows $\Gamma_1$.
 By $i,j \in \Gamma_0$ we label the nodes, and by $e: i \to j$ we denote an arrow $e$ from the source $i = s(e)$ to the target $j = t(e)$. We allow loops and multiple arrows.

\subsection{Cartan matrix and Kac--Moody  algebra}

 A quiver $\Gamma$ defines $|\Gamma_0| \times |\Gamma_0|$ matrix $(c_{ij})_{i,j\in \Gamma_0}$
\begin{equation}
  c_{ij} = 2 - \#(e: i \to j)  - \#(e: j \to i)
\end{equation}
that is called \emph{quiver  Cartan matrix}.
 By definition, the quiver Cartan matrix $c$ is symmetric.
 If there are no single node loops, all diagonal entries of the quiver Cartan matrix are equal to $2$ and such Cartan matrix defines Kac--Moody algebra $\g_\Gamma$ with Dynkin graph $\Gamma$. 

\subsection{Quiver sheaves}

 Choose \emph{the space-time} to be a complex variety $\spacetime$ with structure sheaf $\CalO_\spacetime$, and let $\mathrm{Coh}(\spacetime)$ denote the category of coherent sheaves on $\spacetime$ (the category $\CalO_{\spacetime}\mathbf{Mod}$ of $\CalO_{\spacetime}$-modules).
 Let $\mathrm{Coh}(\spacetime)_{\Gamma} = \mathrm{Rep}(\Gamma, \mathrm{Coh}(\spacetime))$ be the category of representations of quiver  $\Gamma$ in $\mathrm{Coh}(\spacetime)$. We call $\mathrm{Rep}(\Gamma, \mathrm{Coh}(\spacetime))$  by \emph{$\Gamma$-quiver gauge theory on $\spacetime$}: each node $i$ is sent to a sheaf $\CalY_i$ on $\spacetime$ and each arrow $e: i \to j$ is sent to an element of $\Hom_{\CalO_{\spacetime}}(\CalY_i, \CalY_j)$.
 
 In the context of $\CalN=2$ gauge theories, a sheaf $\mathcal{Y}_i$ represents gauge connection in the $i$-th \emph{vector multiplet}, and an element  in $\Hom_{\CalO_{\spacetime}}(\CalY_i, \CalY_j)$ represents field in the $i \to j$ \emph{bi-fundamental} hypermultiplet. 

\subsection{Moduli space}

 Let 
\begin{equation}
  \mathfrak{M}(\Gamma, \spacetime) =   \mathrm{Coh}(\spacetime)_{\Gamma}
/\mathrm{Aut}  (\mathrm{Coh}(\spacetime)_{\Gamma})
\end{equation}
be the moduli space of $\Gamma$-quiver sheaves on $\spacetime$. Let $\gamma = \ch \CalY$ denote the Chern character of the collection $\CalY = (\CalY_i)_{i \in \Gamma_0}$ so that $\gamma = (\gamma_i)_{i \in \Gamma_0}$ with $\gamma_i = \ch \CalY_i \in H^{\bullet}(\spacetime)$.
 The Chern character $\gamma_i$ characterizes the topological class of sheaf $\mathcal{Y}_i$. 

The total moduli space $ \mathfrak{M}(\Gamma, \spacetime)$ of $\Gamma$-quiver sheaves on $\spacetime$ is a disjoint union over topological sectors
\begin{equation}
  \mathfrak{M}(\Gamma, \spacetime)  =  \coprod_{\gamma}   \mathfrak{M}(\Gamma, \spacetime) _{\gamma}
\end{equation}
Algebraically, the moduli space $ \mathfrak{M}(\Gamma, \spacetime)$ is a derived stack with virtual tangent bundle $T\mathfrak{M}(\Gamma, \spacetime)$ at $\CalY$ given by 
\begin{equation}
T_{\CalY} \mathfrak{M}(\Gamma, \spacetime) = \mathrm{Coh}(\spacetime)_{\Gamma}(\CalY, \CalY)[1]
\end{equation}
More explicitly, 
\begin{equation}
\label{eq:TM}
  T_{\CalY} \mathfrak{M}(\Gamma, \spacetime)^{\bullet} = \bigoplus_{ (i
  \stackrel{e}{\to} j) \in \Gamma_1} \Ext_{\CalO_\spacetime}^{\bullet}(\CalY_i, \CalY_j) 
\oplus \bigoplus_{i \in \Gamma_0} \Ext_{\CalO_\spacetime}^{\bullet +1}(\CalY_i, \CalY_i)
\end{equation}

\subsection{Universal sheaf}
Let $\CalY = (\hat{\CalY}_i)_{i \in \Gamma_0}$ denote \emph{the universal sheaf} over $\mathfrak{M}(\Gamma, \spacetime) \times \spacetime$ that is associated to the family of sheaves $\CalY_i$ on $\spacetime$ parametrized by $ \mathfrak{M}(\Gamma, \spacetime) $.

\subsection{Equivariant version}

 Suppose we are given an equivariant action of a complex group $\sT$ on the sheaves $\mathrm{Coh}(\spacetime)$.
 Then quiver gauge theory can be defined $\sT$-equivariantly.
 In particular, group $\sT$ acts on the moduli space $\mathfrak{M}(\Gamma, \spacetime)$ of $\sT$-equivariant $\Gamma$-quiver sheaves on $\spacetime$.

\subsection{Partition function}

 Define \emph{partition function} $Z_{\sT}(\Gamma,\spacetime)_{\gamma}$ in topological sector $\gamma$ be the $\sT$-equivariant index (holomorphic equivariant Euler characteristic) of the structure sheaf on the moduli space of $\Gamma$-quiver sheaves on $\spacetime$ of charge $\gamma$
\begin{equation}
  Z_{\sT}(\Gamma,\spacetime)_\gamma = \sum_{n \in \BZ} (-1)^{n} \ch_{\sT}
H^{n}(\mathfrak{M}(\Gamma, \spacetime)_\gamma, \CalO_{\mathfrak{M}(\Gamma, \spacetime)_\gamma})
\end{equation}
 The total partition function is the sum over the charges 
\begin{equation}
  Z_{\sT}(\Gamma,\spacetime)= \sum_{\gamma} \fq^{\gamma}   Z_{\sT}(\Gamma,\spacetime)_{\gamma}
\end{equation}
 This partition function in the context of $\mathcal{N}=2$ gauge theories is known under the name \emph{K-theoretic} Nekrasov partition function, or the partition function of the 5d quiver gauge theory reduced on $S^1$ \cite{Nekrasov:2002qd} where $\fq$ is a union of $\fq_i = \exp\left( 2\pi \imath \tau_i \right)$ with the complexified coupling constant $(\tau_i)_{i \in \Gamma_0}$.

We can write the partition function using the notation of the derived pushforward
$\pi_{!} = \sum (-1)^{i} R_i \pi_{*}$ for the projection (integration)
map   $  \pi:  \mathfrak{M}(\Gamma, \spacetime) \to
\mathrm{point}$ 
\begin{equation}
\label{eq:Zdef}
  Z_{\sT}(\Gamma, \spacetime) = \ch_{\sT} \pi_! \fq^{\gamma}
\end{equation}
 By definition, the partition function $Z_{\sT, \gamma}$, being a character of a virtual representation in $\mathrm{Rep}(\sT)$, can be evaluated on an element $\st \in \sT$.
 In the context of Nekrasov's partition function the element $\st$ comprises all \emph{equivariant parameters}.

\subsection{Fundamental matter}

 Since quiver $\Gamma$ is arbitrary, unlike \cite{Nekrasov:2013xda,Nekrasov:2012xe} where $\Gamma$ was of finite or affine type, in the present formalism  (anti) fundamental matter for a node $i$ is treated simply as bi-fundamental arrow between the node $i$ and another \emph{frozen node}, denoted by $i'$, which is represented in constant sheaves with gauge coupling constant $\fq_{i'}$ turned off.

\subsection{Local observables}
 Let  $o \in \spacetime$ be a $\sT$-invariant point on space-time $\spacetime$ and let $i_{o}: o \to \spacetime$ be the inclusion map that naturally induces $i_{o}: \mathfrak{M}(\Gamma, \spacetime)  \to \mathfrak{M}(\Gamma, \spacetime) \times \spacetime$.  

 We define \emph{observable sheaves} $(\bY_i)_{i \in \Gamma_0}$ over the moduli space $\mathfrak{M}(\Gamma, \spacetime)$ as the pullback of the universal sheaf $(\hat \CalY_i)_{i \in \Gamma_0}$ from $\mathfrak{M}(\Gamma, \spacetime) \times \spacetime$ to $\mathfrak{M}(\Gamma, \spacetime)$ by the inclusion $i_{o}$
\begin{equation}
  \bY_i = i_o^{*} \hat \CalY_i 
\end{equation}
 Let $\bY^{[p]}_i$ be the $p$-th Adams operation applied to $\bY_i$, which is realized using the complex scalar in the vector multiplet as $\Tr \Phi_i^p$ in 4d.
 The sheaves $(\bY^{[p]}_i)_{i \in \Gamma_0, p \in \BZ_{\geq 1}}$ generate the \emph{ring of observables} (using the direct sum modulo equivalence from exact sequences as the addition and the tensor product as the multiplication) which is a subring in the $\sT$-equivariant K-theory of sheaves on $\mathfrak{M}(\Gamma, \spacetime)$. 

\subsection{Extended partition function}

 We fix a quiver $\Gamma$ and the space-time $\spacetime$ and drop the symbols from the notations. 

 Associated to the local observables $(\bY^{[p]})_{i \in \Gamma_0, p \in \BZ_{\geq 1}}$, introduce parameters, called \emph{higher times} $t = (t_{i,p})_{i \in \Gamma_0, p \in \BZ_{ \geq 1}}$ and \emph{Chern--Simons levels} $(\kappa_i)_{i \in \Gamma_0}$.
 We treat higher times  $t_{i,p}$ as the \emph{conjugate} variable to $\bY_i^{[p]}$ in the sense of the \emph{generating  function}~\cite{Marshakov:2006ii}
 \begin{equation}
\label{eq:generating-function}
 Z_{\sT}(t)
 = \ch_{\sT} \pi_{!} \fq^{\gamma}  \prod_{i \in \Gamma_0}
 [\det \hat \bY_i]^{\kappa_i} \exp\left( \sum_{p = 1}^{\infty} t_{i,p} \bY_{i}^{[p]} \right)
 \end{equation}

 \subsection{Localization}

 Suppose that space of the $\sT$-fixed points in $\frakM$ is a
discrete set of points $\frakM^{\sT}$ with inclusion $i_{\sT}:
\frakM^{\sT} \hookrightarrow \frakM$.
 Then the generating function (\ref{eq:generating-function}) can be computed by \emph{localization
  formula}\footnote{Lefshetz---Grothendieck--Hirzebruch--Riemann--Roch---Atiyah--Singer formula}:
\begin{equation}
\label{eq:ZlocT}
  Z_{\sT}(t)  = \sum_{\frak{M}^{\sT}} \fq^{\gamma} \exp{\left(\sum_{p=1}^{\infty} \frac{1}{p} \ch_{\sT}
      (T^{*}_{\frakM^{\sT}} \frakM )^{[p]} \right)}\prod_{i
        \in \Gamma_0} \ch_{\sT} [\det i^*_{\sT} \hat \bY]^{\kappa_i} \exp{\left( \sum_{p=1}^{\infty} t_{i,p} \ch_{\sT}
        (i^{*}_{\sT}\bY_{i})^{[p]}\right)}
\end{equation}

\section{Quiver gauge theory on $\BC^2$}

 In this section we specialize to the space-time $\spacetime = \BC^2$ with marked point $o \in \spacetime$ and the natural action of complex group $GL(2)$ on $\spacetime$ by with fixed point $o$. 

\subsection{Automorphism group $GL(\bQ)$}

 We denote by $\bQ$ the fiber of the cotangent bundle to $\spacetime$ at $o$
\begin{equation}
  \bQ = T^{*}_{o} \spacetime
\end{equation}
 Then $\bQ$ is the defining module for the group of its automorphisms $GL(\bQ)$.
 We split $\bQ = \bQ_1 \oplus \bQ_2$ with respect to Cartan torus  
 \begin{equation}
\sT_{\bQ}  \simeq  GL(\bQ_1) \times GL(\bQ_2) \subset GL(\bQ)
 \end{equation}
and define $q_1, q_2$ to be corresponding characters 
 \begin{equation}
   q_1 = \ch \bQ_1, \quad q_2 = \ch \bQ_2, \quad Q = \ch_{\sT_{\bQ}} \bQ =
   q_1 + q_2, \qquad q = q_1 q_2 = \ch \Lambda^{2} \mathbf{Q} 
 \end{equation}
\begin{remark*}
 The parameters $(q_1,q_2)$ are exponentiated (multiplicative) \emph{$\ep$-parameters} of the gauge theory~\cite{Moore:1997dj, Nekrasov:2012xe}.
\end{remark*}

\subsection{Automorphism groups  $GL(\bN)$ and $GL(\bM)$}
 Let $\CalO(\spacetime) = \BC[z_1, z_2]$, the ring of polynomials in two variables, be the coordinate ring of $\spacetime = \BC^2$. 
 The space of sections of a coherent sheaf on  $\spacetime = \BC^2$ is a $\CalO(\spacetime)$-module.
 Then $\Gamma$-quiver gauge theory on $\spacetime$ is identified with representation of  $\Gamma$ in $\CalO(\spacetime)$-modules.
 We can take $\BC^2 \simeq \BP^2 \setminus \BP^1_{\infty}$ and fix framing at the $\BP^{1}_{\infty}$. 

 For a $\Gamma$-quiver sheaf $\CalY$, let $\sn \in \BZ_{>0}^{\Gamma_0}$ denote the  rank 
and let $\sk \in \BZ_{\geq 0}^{\Gamma_0}$ denote the instanton charge 
\begin{equation}
 \sn (\CalY) =  \ch_0 \CalY,   \qquad  \sk (\CalY) =- \ch_2 \CalY
\end{equation}
 Let $\bN = (\bN_i)_{i \in \Gamma_0}$ be the framing space associated to $\Gamma_0$ part of quiver (nodes).
To each node $i$ we associate the framing $\bN_i \simeq \BC^{\bn_i}$
for the respective sheaf $\CalY_i$ on $\spacetime$.  Let $GL(\bN) = \prod_{i \in
  \Gamma_0} GL(\bN_i)$ be the respective  group of automorphisms and let
$\sT_{\bN}$ be a Cartan torus of $GL(\bN)$. Let $N_i$ be the
character of $\bN_i$
\begin{equation}
  N_i = \ch_{\sT_{\bN}} \bN_i = \nu_{i,1} + \dots + \nu_{i,\sn_i}
\end{equation}
\begin{remark*}
 The parameters $(\nu_{i,\alpha})_{i \in \Gamma_0, \alpha \in [1\dots \sn_i]}$ are the multiplicative \emph{Coulomb parameters} of the gauge theory. 
\end{remark*}

 Let $\bM = (\bM_e)_{e \in \Gamma_1}$ be the framing space associated to $\Gamma_1$ part of quiver (arrows).
 To each individual arrow $e$ we associate one-dimensional mass-twisting space $\bM_e \simeq \BC$.
 Let $GL(\bM) = \prod_{e \in \Gamma_1} GL(\bM_{e})$ be the respective group of automorphisms and let $\sT_{\bM}$ be a Cartan torus in $GL(\bM)$.
 Let $M$ be the character of $\bM$ 
\begin{equation}
  M = \ch_{\sT_{\bM}} \bM = \mu
\end{equation}
\begin{remark*}
 The parameters $(\mu_{e})_{e \in \Gamma_1}$ are multiplicative \emph{mass parameters} of the bifundamental fields $e: i \to j$ of the gauge theory. 
\end{remark*}

 If we assign a multiplicity $\sm_{e}$ to an arrow $e \in \Gamma_1$, then the mass-twisting space is $\bM_{e} \simeq \BC^{\sm_e}$ with the character
\begin{equation}
  M = \ch_{\sT_{\bM}} \bM = \mu_1 + \dots + \mu_{\sm_e}
\end{equation}
 Since in our formalism $GL(\bM_e)$ is reduced to its Cartan torus $\sT_{\bM_e}  = (\BC^{\times})^{\sm_e}$, the formalism where an arrow  $e: i \to j$ is assigned a multiplicity $\sm_{e}$ is equivalent to the formalism where this arrow is replaced by $\sm_{e}$ for individual arrows $i \to j$. 

\subsection{Complete group of equivariance}

 For $\Gamma$-quiver gauge theory on $\spacetime  = \BC^2$ we denote by
\begin{equation}
  \sT = \sT_{\bQ} \times \sT_{\bN} \times \sT_{\bM}
\end{equation}
the Cartan torus in the automorphism group of the moduli space
$\frakM(\spacetime, \Gamma)$.

\subsection{Fundamental matter as background of higher times}

 Alternatively, fundamental matter can be realized as a background in higher times.
 To add a fundamental hypermultiplet with multiplicative mass $\mu \in \BC^\times$ to the node $i$ it is sufficient to additively modify the times to 
\begin{equation}
 t_{i,p} \to t_{i,p} +  \frac{1}{p} \frac{q^p}{(1 -q_1^p)(1-q_2^p)} \mu^{-p}
  \label{eq:t-matter-shift}
\end{equation}
 To simplify presentation we don't keep track of the fundamental matter since it is a particular case of the higher times theory. 
 In the operator formalism, on the other hand, this shift is imposed by additional vertex operators introduced in Sec.~\ref{sec:V-op}.

\subsection{Localization in quiver theory on  $\BC^2$}
 The localization formula (\ref{eq:ZlocT}) to the $\sT$-fixed point set $\frakM^{\sT}$ in the moduli space of $\Gamma$-quiver sheaves on $\spacetime = \BC^2$ can be explicitly computed~\cite{Moore:1997dj, nakajima-hilbert, Nekrasov:2002qd}. 

 The $\sT$-fixed sheaves split into the direct sum of one-dimensional $\sT$-fixed ideal sheaves, which are classified as $\sT_{\bQ}$-fixed ideals in $\CalO(\spacetime) \simeq \BC[z_1, z_2]$ where the fixed point $o \in \BC^2$ is the origin $o = (0,0)$. A $\sT_{\bQ}$-fixed ideal in $\BC[z_1, z_2]$ of $\ch_2 = - \sk$ is labelled by a partition $(\lambda) = \lambda_1 \geq \lambda_2 \geq \dots \geq 0  \geq 0 \dots $ of total size $|\lambda| = \sum_{i=1}^{\infty} \lambda_i  = \sk$.
 Each box $s = (s_1, s_2)$ in the partition $\lambda$ with $s_1 \in [1 \dots \infty]$ and $s_2 \in [1 \dots \lambda_{s_1}]$  is associated to the monomial $z_1^{s_1 - 1} z_2^{s_2 - 1}$.
 The ideal $\bI_{\lambda} \subset \CalO(\spacetime) = \bI_{\emptyset}$ is $\CalO(\spacetime)$-generated by all monomials outside of the partition $\lambda$. 
 Let $\bK_{\lambda} = \bI_{\emptyset}/\bI_{\lambda}$ be generated by the monomials in the partition $\lambda$. 

A $\sT_{\bN} \times \sT_{\bQ}$-fixed $\CalO(\spacetime)$-module $ \bY_{\spacetime}
=\hat \CalY(\spacetime)$ of rank $\sn$
splits into direct sum of  $\sT_{\bQ}$-fixed ideals
\begin{equation}
  \bY_{\spacetime} = \bigoplus_{\alpha \in [1 \dots \sn]} \bI_{\lambda_\alpha}
  \otimes \bN_{\alpha}
\end{equation}
and let $\bY \equiv \bY_{o} = i_{o}^{*}  \bY_{\spacetime}$. Then we have in K-theory by
localization to $i_{o}: o \hookrightarrow \spacetime$ 
\begin{equation}
  [\bY_{\spacetime} ]=[ \bY_o] / [\Lambda \bQ]
\end{equation}
where $\Lambda \bQ = \sum_{i} (-1)^{i} \Lambda^{i} \bQ$
and the division is in formal series. Then 
\begin{equation}
 [\bY_o] = [\bN] - [\Lambda \bQ] [\bK ]
\end{equation}

\subsection{Cotangent moduli space}
From (\ref{eq:TM}) we find the K-theory class  $[T_{\CalY}^{*} \frakM]$ at
$\sT$-fixed point $\hat \bY \in \frakM^{\sT}$
\begin{equation}
\label{eq:TMY}
  [T_{\hat \bY}^{*} \frakM] = \frac{1}{  [\Lambda \bQ^{\vee}]}
 \left (\sum_{(i
  \stackrel{e}{\to} j) \in \Gamma_1 } [\bM_{e}^{\vee}][
  \bY_{o}]_i[\bY_{o}^{\vee}]_j - \sum_{i \in \Gamma_0} [\bY_{o}]_i
  [\bY_{o}^{\vee}]_i \right)
\end{equation}

\subsection{Two commutative reductions}
Since the space-time $\spacetime$ is a product $\spacetime = \spacetime_1 \times
\spacetime_2$ the reduction from $\bY_{\spacetime}$ to $\bY_{o}$ can be done in
two steps in two ways, either first project along $\spacetime_2$ and then
along $\spacetime_1$ (left path) or first project along $\spacetime_1$ and then
along $\spacetime_2$ (right path)
\begin{equation}
  \begin{tikzcd}
{}   &  {}{[\bY_{\spacetime}]} \arrow{dd}{\cdot [\Lambda \bQ] } \arrow{dl}[swap]{\cdot
  [\Lambda \bQ_2]}  \arrow{dr}{\cdot   [\Lambda \bQ_1]}&  {}\\
{}[\bX]:={[\bY_{\spacetime_1}]} \arrow{dr}[swap]{\cdot [\Lambda \bQ_1]}  &
{} &{}{[\bY_{\spacetime_2}] }\arrow{dl}{\cdot [\Lambda \bQ_2]}=:[\tilde \bX] \\
{}  &   {}{[\bY_{o}]}  & {} 
  \end{tikzcd}
\end{equation}
so that it holds 
\begin{equation}
 [\bY_{o}] = [\Lambda \bQ_1][\bX], \qquad
 [\bY_{o}] = [\Lambda \bQ_2] [\tilde \bX]
\end{equation}
\begin{remark*}
 Swapping $\spacetime_1 \leftrightarrow \spacetime_2$ leads to transposition of the partition $(\lambda_{i,\alpha}) \leftrightarrow (\lambda_{i,\alpha}^\text{T})$ characterizing the $\sT$-fixed point in the moduli space.
\end{remark*}

\subsection{Quantum $q$-geometric Langlands Duality}
The exchange  $\spacetime_1 \leftrightarrow \spacetime_2$ in the above diagram leads to the
\emph{quantum q-geometric Langlands duality} $q_1 \leftrightarrow q_2$.
See section~\ref{se:Introduction} for references.

\subsection{Intermediate reduction}
 The class $[T^{*}_{\hat \bY} \frakM]$ at fixed point $\hat \bY \in
 \frakM^{\sT}$  in the equation (\ref{eq:TMY}) of the
partition function can be expressed in terms of $[\bX]\equiv [\bY_{\spacetime_1}]$ 
\begin{equation}
 [ T^{*}_{\bY} \frakM] = \frac{ [\Lambda \bQ_1]}{ [\Lambda \bQ_2^{\vee}]} 
  \left (   - \sum_{(i,j) \in \Gamma_0 \times \Gamma_0}
   [\bX]_i c_{ij}^{+} [\bX]_j^{\vee}  \right)   
\end{equation}
and the K-theory valued \emph{half Cartan matrix} $c_{ij}^{+}$
defined as 
\begin{equation}
 [\bc_{ij}^{+}] :=
  \delta_{ij} - \sum_{e:i \to j} [\bM^{\vee}_{e}]  
\end{equation}
with Chern character
\begin{equation}
 \ch [\bc_{ij}^{+}] = \delta_{ij} - \sum_{e:i \to j} \mu_e^{-1}  
\end{equation}

\subsection{The set of eigenvalues}
 Then the Chern characters  $X = \ch \bX$ at $\sT$-fixed point $\lambda$  can be explicitly described.
 Let
\begin{equation}
 \CalX_i= \{ x_{i,\alpha,s_1} \}_{ \alpha \in [1\dots \sn_i], s_1 \in [1 \dots \infty] }, \qquad
 \CalX = \bigsqcup_{i \in \Gamma_0} \CalX_i
\end{equation}
be the set of characters of the monomials associated to boxes
$(s_1, \lambda_{s_1} + 1)$ that generate $(\bY_{\spacetime})$ as
$\CalO(\spacetime_{2})$-module so that 
\begin{equation}
x_{i, \alpha, s_1} = \nu_{i,\alpha} q_1^{s_1-1}
q_2^{\lambda_{i,\alpha,s_1}}
\end{equation}
and $ X_{i} = \sum_{x \in \CalX_i} x$. 
Let $\si: \CalX \to \Gamma_0$ be the node label so that $\si(x)=i$ for $x
\in \CalX_i$.

\subsection{The partition function}

 In terms of $x$-variables the extended partition function (\ref{eq:ZlocT}) is 
\begin{multline}
\label{eq:partx}
  Z_{\sT}(t) = \sum_{\CalX \in \frakM^{\sT}} \exp\left( - \sum_{ (x_L,x_R)\in \CalX
        \times \CalX} \sum_{p=1}^{\infty}\frac{1}{p}
    \frac{1 - q_1^{p}}{1 - q_2^{-p}} (c^{+}_{\si(x_L), \si(x_R)})^{[p]} x_L^{-p} x_R^{p}
    \right) \\\times 
\exp\left( \sum_{x \in \CalX}  \left( - \frac {\kappa_{\si(x)}} 2 ( \log_{q_2} x -1) \log_{q_2} x+  \log \fq_{\si(x)} \log_{q_2} \frac{x}{\mathring{x}} + \sum_{p=1}^{\infty} (1
  -q_1^{p}) t_{\si(x),p} x^{p} \right) \right)
\end{multline}
where $\mathring{x}_{i, \alpha, s_1} = \nu_{i,\alpha} q_1^{s_1 -1}$ denotes \emph{ground configuration} of the empty partition $\lambda = 0$, so that the $\log_{q_2} \frac{x}{\mathring{x}}$ counts the size of the partition $\lambda$ equal to the instanton charge $\sk$.
 In particular, the vector and bifundamental hypermultiplet contributions are generated explicitly by the first factor in (\ref{eq:partx}) and are given by
\begin{align}
 Z_i^\text{vec}
 & =
 \prod_{(x,x') \in \CalX_i \times \CalX_i}
 \left( q \frac{x}{x'};q_2 \right)_\infty
 \left( q_2 \frac{x}{x'};q_2 \right)_\infty^{-1}
 \, , \\
 Z_{e:i \to j}^\text{bf}
 & =
 \prod_{(x,x') \in \CalX_i \times \CalX_j}
 \left( \mu_e^{-1} q \frac{x}{x'};q_2 \right)_\infty^{-1}
 \left( \mu_e^{-1} q_2 \frac{x}{x'};q_2 \right)_\infty
 \, ,
\end{align}
These factors correspond to the localization determinants in the full
Nekrasov partition function in quiver gauge theory, including the perturbative one-loop factor (see for example \cite{Nekrasov:2013xda}).
The $\kappa$-term in the second factor in (\ref{eq:partx}) accounts for the Chern--Simons contribution,
the $\fq$-term in the second factor in (\ref{eq:partx}) accounts for the instanton number counting parameter,
and the $t$-terms in the second factor in (\ref{eq:partx}) account for the deformation of the Nekrasov partition function
by the higher times.
\begin{remark*}
 In the limit $q_2 \to 1$ the extended partition function is dominated by the critical set $\CalX_{\text{crit}}$ determined in~\cite{Nekrasov:2013xda} and the variables $x \in \CalX_{\text{crit}}$ satisfy the Bethe equations.
\end{remark*}

\subsection{Reflection of the index}
Let $\bX^{[p]}$ be $p$-th Adams operation applied to an object
$\bX$. In terms of the decomposition of $\bX$ into the Chern roots $\bX=\sum x$,
the $p$-th Adams operation takes each Chern root $x$ into the $p$-th power, so that $\bX^{[p]} = \sum x^p$.
Then the following \emph{reflection equation} holds
\begin{equation}
\label{eq:refl}
 \exp\left( \sum_{p=1}^{\infty} \frac{1}{p} \left[\left(\bX^{\vee}\right)^{[p]}\right] \right)
 =
 (-1)^{\rk \bX} \left[ \det \bX \right]
 \exp\left( \sum_{p=1}^{\infty} \frac{1}{p} \left[ \bX^{[p]} \right] \right)
\end{equation}

\subsection{The ordered partition function}
 Pick an order $\succ$ on the set $\CalX$.
 For example, an order can be chosen  by taking $|q_1| \ll |q_2^{-1}| < 1$ and $|\nu| \simeq |\mu| \simeq 1$.
 Then define $x_{L} \succ x_{R}$ if $ |x_L| > |x_R|$, which is basically the radial order in CFT.
 The sum over all pairs $(x_L, x_R) \in \CalX \times \CalX$ in the partition function (\ref{eq:partx}) can be transformed to the sum over pairs $(x_L \succ x_R)$, over pairs $(x_L \prec x_R)$ and the diagonal pairs $(x_L =x_R).$
 The diagonal part gives $(\fq, \nu, \mu, t)$-independent factor that we omit.
 The sum over pairs $(x_L \succ x_R)$ and $(x_L \prec x_R)$ can be combined together using the reflection equation (\ref{eq:refl})
\begin{multline}
\label{eq:partxorderd}
  Z_{\sT}(t) = \sum_{\CalX \in \frakM^{\sT}} \exp\left(  \sum_{ (x_L \succ x_R) \in \Lambda^2\CalX
   } -(c_{\si(x_R),\si(x_L)}^{+})^{[0]}\beta  \log\frac{x_{R}}{ x_L} -\sum_{p=1}^{\infty}\frac{1}{p}
    \frac{1 - q_1^{p}}{1 - q_2^{-p}} (c_{\si(x_L), \si(x_R)})^{[p]} \frac{ x_R^{p}}{{x_L}^{p}}
    \right) \\\times 
\exp \left( \sum_{x \in \CalX} \left( - \frac {\kappa_{\si(x)}} 2 (
    \log_{q_2} x -1) \log_{q_2} x +  \log \fq_{\si(x)}  \log_{q_2} \frac{x}{\mathring{x}} + \sum_{p=1}^{\infty} (1
  -q_1^{p}) t_{\si(x),p} x^{p}  \right) \right)
\end{multline}
where $\beta = -\frac{\log q_1}{\log q_2} = -\frac{\ep_1}{\ep_2}$, and the \emph{mass deformed} Cartan matrix is defined
\begin{equation}
\label{eq:massC}
 c_{ij} = c_{ij}^{+} + c_{ij}^{-}, \qquad c_{ij}^- = q^{-1}
 (c_{ji}^{+})^\vee, \qquad 
 c_{ij} =
 \left( 1 + q^{-1} \right) \delta_{ij}
 - \sum_{e:i \to j} \mu_e^{-1}
 - \sum_{e:j \to i} \mu_e q^{-1}
\end{equation}
which is symmetric up to conjugation and the $q$-factor,
\begin{align}
 c_{ji} = q^{-1} c_{ij}^\vee
\end{align}

\subsection{Extended partition function is a state}
 The exponentiated sum over the pairs $(x_L \succ x_{R})$  in the equation (\ref{eq:partxorderd}) suggests a natural way to present the extended partition function $Z_{\sT}(t)$ as a state $|Z_{\sT} \rangle$ in the infinite-dimensional $\sT$-character valued Fock space $\ch \mathrm{Rep}_{\sT}[[t]]$.
 The Fock space  $\ch \mathrm{Rep}_{\sT}[[t]]$ is Verma module for the Heisenberg algebra $\bH$ generated by the operators $(\p_{i,p})_{i \in \Gamma_0, p \in [0\dots \infty]}$ and $t = (t_{i,p})_{i \in \Gamma_0, p \in [0 \dots \infty]}$ over $\ch \mathrm{Rep}_{\sT}$ with canonical commutators
\begin{equation}
[\p_{i,m}, t_{j,n}] =\delta_{ij} \delta_{mn}
\end{equation}
where 
\begin{equation}
  t_{i,0} = \log_{q_2} \fq_i 
\end{equation}
 The elements of the Fock space $\ch \mathrm{Rep}_{\sT}[[t]]$ are formal $t$-series valued in the ring of $\sT$-characters. The $t$-constants are lowest-weight states (vacua); they are annihilated by all lowering operators $\p_{i,p}$.
 A state in the Fock space $\ch \mathrm{Rep}_{\sT}[[t]]$ can be obtained by an action of an operator in the algebra $\bH$ on the vacuum $|1\rangle$.

\subsection{Free bosons and  vertex operators}
The state $|Z_{\sT}\rangle$ can be presented as 
\begin{equation}
\label{eq:Sproduct}
  |Z_\sT \rangle =  \sum_{\CalX \in \frakM^{\sT}} \prod^{\succ}_{x \in
    \CalX} S_{\si(x),x}  |
 1 \rangle
\end{equation}
where $\displaystyle \prod^{\succ}$ denotes the $\succ$-ordered product over $x \in \CalX$ of the \emph{vertex operators}
\begin{equation}
\label{eq:S_i}
 S_{i,x} = \
 :  \exp\big(  \sum_{p > 0 } s_{i,-p} x^{p} + s_{i,0} \log x + \tilde s_{i,0} +
  \sum_{p > 0 } s_{i,p} x^{-p}\big):
\end{equation}
Here  the \emph{free field modes} or \emph{oscillators} are
\begin{equation}
    s_{i,-p} \stackrel{p>0}{=} ( 1- q_1^{p}) t_{i,p}, \qquad  s_{i,0} =
    t_{i,0}, \quad 
    s_{i, p} \stackrel{p>0}{=} - \frac{1}{p} \frac{1}{1 - q_2^{-p}}
    c_{ji}^{[p]} \p_{j,p}    
\end{equation}
with commutation relations 
\begin{equation}
\label{eq:scomp}
 [ s_{i,p}, s_{j,p'}] =- \delta_{p+p',0}\frac{1}{p} \frac{1 - q_1^p}{1 -
 q_2^{-p}}
 c_{ji}^{[p]} \qquad \left( p > 0 \right)
\end{equation}
The conjugate \emph{zero mode} $\tilde s_{i,0}$ satisfies
\begin{equation}
 \label{eq:scom} [ \tilde s_{i,0} , s_{j,p}]
  = - \beta  \delta_{0,p} c_{ji}^{[0]} 
\end{equation}
The normal product notation $:e^{A_1} e^{A_2} :$, where operators $A_1,A_2$ are linear in the free fields, means that  all operators $(s_{i,p})_{p\leq 0}$ are placed to the left of $(s_{i,p})_{p>0}$ and $\tilde s_{i,p}$.  

 The relations (\ref{eq:scom}) and (\ref{eq:scomp}), and the relation $e^{A_1} e^{A_2} = e^{[A_1, A_2]} e^{A_2} e^{A_1}$ for central $[A_1,A_2]$, imply that the  operator-state representation of the partition function (\ref{eq:Sproduct}) is equivalent to the quiver gauge theory definition (\ref{eq:partxorderd}) if gauge theory couplings $\kappa_i$ and $\fq_i$ are evaluated as
\begin{equation}
  \begin{aligned}
   \kappa_i & =
   - \sn_j (c_{ji}^{-})^{[0]} , \qquad
   \log_{q_2} \fq_{i} & = \beta + t_{i,0} +
   \sn_j (c_{ji}^{-})^{[\log_{q_2}]}
   - \log_{q_2} ((-1)^{\sn_j} \nu_j) (c_{ji}^-)^{[0]} 
  \end{aligned}
\end{equation}
where 
\begin{equation}
 (c_{ij}^{-})^{[\log_{q_2}]} = \delta_{ij} \log_{q_2} q^{-1}
  - \sum_{e: j \to i} \log_{q_2} ( \mu_{e} q^{-1} )
\end{equation}

\subsection{Screening charges}
 The configuration sets $\CalX \in \frakM^{\sT}$ are described by the partitions, which are explicitly collections of constrained sequences
\begin{equation}
\label{eq:constraints}
(\lambda_{i, \alpha, 1} \geq \lambda_{i, \alpha, 2}
\geq \dots \geq 0  =  0 = 0= \dots )_{i \in \Gamma_0, \alpha \in
  [1\dots \sn_i]}
\end{equation}
Let $\CalX_{0,i}$ be the ground configuration with all $\lambda_{i, \alpha, *} = 0$, and define its union $\CalX_{0} = \bigsqcup_{i \in \Gamma_0} \CalX_{0,i}$.
 Let $\BZ^{\CalX_{0,i}}$ be the set of collections of arbitrary integer sequences terminating by zeroes 
\begin{equation}
(\lambda_{i, \alpha, s_1} \;? \; \lambda_{i, \alpha, s_2} \; ?
  \dots \; ?  = 0 = 0 = \dots )_{i \in \Gamma_0, \alpha \in
  [1\dots \sn_i]}
\end{equation}
 Then $\frakM^{\sT} \subset \BZ^{\CalX_0}$.
 It turns out that the summation over $\frakM^{\sT}$ in (\ref{eq:Sproduct}) can be extended to the whole $\BZ_{\geq 0}^{\CalX_0}$ without changing the result because 
\begin{equation}
   \prod^{\succ}_{x \in \CalX} S_{\si(x),x} | 1 \rangle = 0 \qquad
   \text{if $\CalX \in \BZ^{\CalX_0}$ but $\CalX \notin \frakM^{\sT}$}
\end{equation}
due to the zero factors in the normal ordering product of vertex operators $S_{i,x}$ for the sequences $(x_{i, \alpha, s_1} = \nu_{i,\alpha} q_1^{s_1 -1} q_2^{\lambda_{i, \alpha,s_1}})$ where $\lambda$ does not satisfy the constraint (\ref{eq:constraints}). 
Therefore
\begin{equation}
    |Z_\sT \rangle =  \sum_{\CalX \in \BZ^{\CalX_0}} \prod^{\succ}_{x \in
    \CalX} S_{\si(x),x} |
 1 \rangle
\end{equation}

For every point $\mathring{x}_{i, \alpha, s_1} = \nu_{i,\alpha} q_1^{s_1 -1}$ in the ground configuration $\CalX_0$ defines the operator called
\emph{screening charge}
\begin{equation}
\label{eq:St}
  \sS_{i, \mathring{x}} = \sum_{s_2 \in \BZ} S_{i,q_2^{s_2} \mathring{x}}
\end{equation}
Then the state  $|Z_{\sT}\rangle$ is obtained by applying to the vacuum the ordered product of $\sS_i$ operators 
\begin{equation}
\label{eq:ZTSS}
      |Z_\sT \rangle =   \prod^{\succ}_{\mathring{x} \in
    \CalX_0} \sS_{\si(\mathring{x}),\mathring{x}} |
  1 \rangle
\end{equation}
The partition function of plain, not $t$-extended theory, can be interpreted as the  projection
\begin{equation}
  \langle 1 | Z_{\sT} \rangle  =   \langle 1 | \prod^{\succ}_{\mathring{x} \in
    \CalX_0} \sS_{\si(\mathring{x}), \mathring{x}}  | 1 \rangle
    \label{eq:plain-Z-func}
\end{equation}
since the dual vacuum obeys $\bra{1} t_{i,n} = 0$ for $n \ge 1$, $\forall i \in \Gamma_0$.

\subsection{The Ward identities}
\label{eq:Ward}
The sum representation of the operator $\sS_{i, \mathring{x}}$ in (\ref{eq:St}) is explicitly invariant under the
$\BZ$-translational symmetry $s_2 \to s_2 + \BZ$ (change of variables). Hence the
representation of the partition function  (\ref{eq:ZTSS}) is invariant
under the  $\BZ^{\CalX_0}$  symmetry that shifts the summation variables
$s_2$ for each $\mathring{x}_{i,\alpha,s_1}$. In the $\spacetime$ space-time
picture the variation  $s_2 \to s_2 + 1$ amounts to the $z_2$
multiplicative change of variables in the $z_1, z_2$-mode expansion 
 $\phi_{i, \alpha, s_1} z_1^{s_1} \to z_2 \phi_{i, \alpha,   s_1}
 z_1^{s_1}$ where $\phi$ is in the sheaf $\CalY$.  The shift $s_2 \to s_2 + 1$ adds one box to the partition,  or
 equivalently one instanton to the gauge field on the space-time.

\subsection{The Y-operators}

 In \cite{Nekrasov:2012xe, Nekrasov:2013xda} the $(\bY_{i,x})_{i \in \Gamma_0}$ observables were introduced in the K-theory of the moduli space $\frakM^{\sT}$ of the quiver gauge theory
\begin{equation}
  \bY_{i,x} := \exp \left( - \sum_{p=1}^{\infty} \frac{x^{-p}}{p}  \bY_i^{[p]} \right)
\end{equation}
 The expectation value of the observable $\bY_{i,x}$  in the plain (not $t$-extended) theory is computed by the pushforward integration over the moduli space $\frakM^{\sT}$ (\ref{eq:generating-function})
 \begin{equation}
\label{eq:Yexpect}
\langle \bY_{i,x}\rangle   := \ch_{\sT} \pi_{!} \fq^{\gamma} \bY_{i,x}
 \end{equation}

It is natural to lift the $\bY_{i,x}$ observables to the $t$-extended
theory by giving them the operator definition:
\begin{equation}
  \sY_{i,x} = 
 q_1^{\tilde \rho_i}
:\exp\big( \sum_{p > 0} y_{i,-p} x^{p} + y_{i,0} + \sum_{p>0} y_{i,p}
  x^{-p} \big):
\end{equation}
where $\tilde \rho_i := \sum_{j \in \Gamma_0} \tilde c_{ji}^{[0]}$ are components of the Weyl vector in the basis of simple roots.
 If the quiver is the affine type, we put $\tilde \rho_i=0$. %
 The operator $\sY_{i,x}$ is an element of the Heisenberg algebra $\bH$.
 The oscillators $y_{i,p}$ are expressed in terms of $t_{i,p}$ and $\p_{i,p}$
\begin{align}
\label{eq:trel}
 (p>0) \qquad
 y_{i,-p} = (1 - q_1^{p})(1 - q_2^{p})
 \tilde c^{[-p]}_{ji} 
 t_{j,p}, \qquad y_{i,0} =
 - t_{j,0} \tilde c^{[0]}_{ji}
 \log {q_2}  \qquad   y_{i,p} = -\frac{1}{p} \p_{i,p}
\end{align}
or equivalently terms of the free field $s_{i,p}$ 
\begin{equation}
\label{eq:yins}
 y_{i,p} \stackrel{p\neq 0}{=} (1-q_2^{-p}) s_{j,p} \tilde{c}_{ji}^{[p]} , \qquad
 y_{i,0} =(\log q_2^{-1}) s_{j,0} \tilde{c}_{ji}^{[0]}  
\end{equation}
where $\tilde c_{ij}$ is the inverse to the mass-deformed  Cartan matrix
$c_{ij}$ defined in (\ref{eq:massC}). 
 The definitions (\ref{eq:trel}) and (\ref{eq:generating-function}) imply
\begin{equation}
\label{eq:ys}
 \langle \bY_{i,x} \rangle  = \langle 1 | \sY_{i,x} |Z_{\sT}\rangle
\end{equation}

\subsection{The OPE of $\sY$ and $S$}

 The commutation relations between $y_{i,p}$ and $s_{j,p'}$ are
\begin{equation}
\label{eq:ys-commutator}
  [y_{i,p}, s_{j,p'}] = -\frac{1}{p} (1 - q_1^{p})
  \delta_{p+p',0}\delta_{ij}, \qquad [ \tilde s_{i,0}, y_{j,0}] =
  -\delta_{ij} \log q_1
\end{equation}
 Then (\ref{eq:ys}) can be also seen from the  commutation relations (\ref{eq:ys-commutator}) and normal ordering because at $|x| > |x'|$ we have
\begin{equation}
\label{eq:YS1}
  \sY_{i,x}  S_{i,x'}  =\frac{1 - x'/x}{1 - q_1 x'/x} :  \sY_{i,x}
  S_{i,x'}:\qquad \sY_{i,x} S_{j,x'} = \, :\sY_{i,x} S_{j,x'}: \quad i  \neq j
\end{equation}
 Therefore at each fixed point configuration $\CalX \in \frakM^{\sT}$
\begin{equation}
\label{eq:YSstate}
\langle 1 |  \sY_{i,x}  \prod^{\succ}_{x' \in  \CalX} S_{\si(x),x'}  |
1 \rangle =
q_1^{\tilde{\rho}_i}
\left(\prod_{x' \in \CalX_{i}} \frac{ 1-x'/x}{1 - q_1 x'/x}
\right)\langle 1 | \prod^{\succ}_{x' \in  \CalX} S_{\si(x),x'} | 1 \rangle 
\end{equation}
like in the definition that was given in \cite{Nekrasov:2013xda}.
The observable $\sY_{i,x}$ is not regular in the $\BC^{\times}_{x}$ because of the possible poles at points $x = q_1 x'$.

\subsection{The commutator of  $\sY$ and $S$}
 The commutation relations (\ref{eq:ys-commutator}) also imply for $|x'| > |x|$ 
\begin{equation}
\label{eq:SY1}
  S_{i, x'} \sY_{i,x} = q_1^{-1} \frac{1 - x/x'}{1 - q_1^{-1} x/x'} :  S_{i, x'} \sY_{i,x}:
\end{equation}
Therefore (\ref{eq:YS1}) and (\ref{eq:SY1}) imply the non-zero \emph{radial-ordered} commutator 
\begin{equation}
\label{eq:YScommutator}
  [\sY_{i,x}, S_{i,x'}] = (1 - q_1^{-1}) \delta ( q_1 \frac{x'}{x}) :  S_{i, x'} \sY_{i,x}:
\end{equation}
where by definition $ \delta(z) = \sum_{n \in \BZ} z^{n} $. 

 The fact that observable $\sY_{i,x}$ has singularities at $x = q_1 x'$ in (\ref{eq:SY1}) is equivalent to the presence of the $\delta(q_1 x'/x)$ in the radial ordered commutator between $\sY_{i,x}$ and $S_{i,x'}$.
 This is a general statement implied by Cauchy integral formula and familiar from the formalism of radial quantization in CFT.

\subsection{The $\sV$-operators}
\label{sec:V-op}

 We introduce another kind of vertex operator to reproduce the fundamental matter contribution in gauge theory.
 As explained before, this contribution is given by shift of the time variables~\eqref{eq:t-matter-shift}, which can be implemented by the operator
\begin{align}
 \sV_{i,x} & = \
 : \exp
 \left(
 \sum_{p>0} v_{i,-p} x^{p} + \sum_{p>0} v_{i,p} x^{-p}
 \right) :
 \, .
\end{align}
 The corresponding free field is explicitly written
\begin{align}
 (p>0)
 \qquad
 v_{i,-p} =
 - t_{j,p} \tilde{c}^{[-p]}_{ji} 
 \qquad
 v_{i,p} = \frac{1}{p} \frac{1}{(1-q_1^p)(1-q_2^p)} \partial_{i,p}
 \, .
\end{align}
 We remark a simple relation to the $y$-operators \eqref{eq:trel}
\begin{align}
 v_{i,p} & = - \frac{1}{(1-q_1^p)(1-q_2^p)} y_{i,p}
 \, .
\end{align}
 Then the OPE of $\sV$ and $S$ operators are given by
\begin{align}
 \sV_{i,x} S_{i,x'}
 =
 \left(
  \frac{x'}{x};q_2
 \right)_\infty^{-1}
 :\sV_{i,x} S_{i,x'}:
 \, ,
 \qquad
 S_{i,x'} \sV_{i,x} 
 & =
 \left(
  \frac{q_2 x}{x'};q_2
 \right)_\infty
 :\sV_{i,x} S_{i,x'}:
\end{align}
corresponding to the fundamental and antifundamental hypermultiplet contributions.
 Thus the extended partition function in the presence of (anti)fundamental matters is obtained by inserting the $\sV$-operators
\begin{align}
 \ket{Z_\sT} & =
 \left( \prod_{x \in \CalX_\text{f}} \sV_{\si(x),x} \right)
 \left(
 \prod^{\succ}_{\mathring{x} \in \CalX_0}
 \sS_{\si(\mathring{x}),\mathring{x}}
 \right)
 \left( \prod_{x \in \tilde\CalX_\text{f}} \sV_{\si(x),x} \right) 
 \ket{1}
\end{align}
where $\CalX_\text{f} = \{\mu_{i,f}\}_{i\in\Gamma_0,f\in[1\ldots \sn_i^\text{f}]}$ and $\tilde\CalX_\text{f} = \{\tilde\mu_{i,f}\}_{i\in\Gamma_0,f\in[1\ldots \tilde\sn_i^\text{f}]}$ are sets of fundamental and antifundamental mass parameters.
 This $\sV$-operator creates a singularity on the curve at $x = \mu_{i,f}$.
 Then the plain partition function $(t=0)$ is given as a correlator as shown in \eqref{eq:plain-Z-func},
\begin{align}
 Z_\sT(t=0)
 & = \vev{1|Z_\sT}
 =
 \bra{1}
 \left( \prod_{x \in \CalX_\text{f}} \sV_{\si(x),x} \right)
 \left(
 \prod^{\succ}_{\mathring{x} \in \CalX_0}
 \sS_{\si(\mathring{x}),\mathring{x}}
 \right)
 \left( \prod_{x \in \tilde\CalX_\text{f}} \sV_{\si(x),x} \right)  
 \ket{1}
 \, .
\end{align}

\section{W-algebra}

 Here we describe the construction of regular observables $\tT$ of the extended gauge theory and explain isomoprhism with Shiraishi et al.~\cite{Shiraishi:1995rp} and Frenkel--Reshetikhin~\cite{Frenkel:1997} definition of $W_{q_1,q_2}$ algebra as commutant of screening charges in the Heisenberg algebra $\bH$, and define K-theoretical quiver W-algebra for $\spacetime = \BC_{q_1, q_2}$.

\subsection{Pole cancellation in $T$: $A_1$-example}

 Consider the simplest quiver  $\Gamma=A_1$ for example.
 In \cite{Nekrasov:2013xda} in the study of the $q_2 = 1$ limit of the gauge theory partition function, motivated by cut-crossing story of~\cite{Nekrasov:2012xe}, it was suggested to consider the observable%
\footnote{We adopted the normalizations and the zero modes to the conventions of the present paper in which the $\tT$-observables have the simplest canonical form.}  
\begin{equation}
\label{eq:A1-current} T_{1,x} =   \sY_{1,x} + \sY_{1, q^{-1} x}
\end{equation}
for its virtue of being regular function in $\BC^{\times}_{x}$. This is the simplest example of $q$-character representing the $T$-matrix of Baxter coming from $\mathbf{U}_{q}(\hat{ \mathfrak{sl}}_2)$-integrable system and Baxter equation. 

 In fact, the same observable $\tT_{1,x}$  remains regular function of $x$ for generic $q_2$.
 Indeed, in the operator formalism we find 
\begin{equation}
\label{eq:prefactor}
  \begin{aligned}
  \sY_{1,x}  S_{1,x'}  &=\frac{1 - x'/x}{1 - q_1 x'/x} :  \sY_{1,x}
  S_{1,x'}:\\
  \sY^{-1}_{1, q^{-1}x} S_{1, x''} &= \frac{ 1  - q q_1 x''/x}{1 - q x''/x}
  : \sY^{-1}_{1, q^{-1}x} S_{1, x''}:
  \end{aligned}
\end{equation}
so the potential singularity in the first line is for $x' = q_1^{-1}x $ and in the second line for $x'' = q^{-1} x$.
 Therefore, the two singularities have chance to cancel at $x' = q_2 x''$. Recall that the state $|Z_{\sT} \rangle$ is obtained with the sums (\ref{eq:St}) and there is internal symmetry for the shift of the summation indexing  variable (see Ward identity in Sec.~\ref{eq:Ward}) so that for every term $S_{1,x''}$ there is a term with $S_{1,x'}$ with $x' = q_2 x''$. 

Indeed, we find for the first term the normal ordered expression
\begin{equation}
  \begin{aligned}
 : \sY_{1,x} S_{1,q_1^{-1} x}: \, = \, : \exp\left(  - \frac 1 2 s_{1,0} \log
   q_2  +  s_{1,0} \log (x q_1^{-1})
 + \tilde s_{1,0} +  \sum_{p \neq 0} (q_1^p  + \frac{ 1 - q_2^{-p}}{ 1
   + q^{-p}}) s_{1,p} x^{-p}\right):
  \end{aligned}
\end{equation}
and for the second term the normal ordered expression 
\begin{equation}
 q_1 : \sY_{1,q^{-1} x}^{-1} S_{1,q^{-1} x}: \, = \, :\exp\left(   + \frac 1 2
   s_{1,0} \log q_2  + s_{1,0} \log (x q^{-1})
+ \tilde s_{1,0} + \sum_{p\neq0} (q^p - \frac{1 -
     q_2^{-p}}{1 + q^{-p}} q^p)  s_{1,p} x^{-p}
 \right): 
\end{equation}
which are exactly identical.
 The respective residues in the prefactors
(\ref{eq:prefactor}) are $(1-q_1^{-1})$ and $(q_1^{-1} -1)$ which
respectively cancel each other. 
 This computation proves regularity in $x \in \BC^{\times}$ of the state $|Z_{\sT}\rangle $ of higher $t$-extended gauge theory in $A_1$ example
\begin{equation}
\label{eq:Treg}
 \partial_{\bar x}  \tT_{1,x} |Z_{\sT} \rangle = 0
\end{equation}

\subsection{Commutator of  $\tT$ and $\sS$ vanishing: $A_1$-example}

An exactly equivalent presentation of the regularity  of
$\tT_{1,x}$ is the statement that 
\begin{equation}
  [\tT_{1,x}, \sS_{1,x'}] = 0
\end{equation}
where $\sS_{1,x'}$ is the screening charge (\ref{eq:St}) defined as the summation over the $q_2^{\BZ}$ shifts. 
Indeed, we have
\begin{equation}
  \begin{aligned}
{}  [ \sY_{1,x} ,   S_{1, x'}] &=  (1 - q_1^{-1}) \delta ( q_1
  \frac{x'}{x}) :  \sY_{1,x} S_{1, x'} :\\
[ \sY^{-1}_{1, q^{-1} x} ,  S_{1, x'}]  &= (q_1^{-1}
-1) \delta (q \frac{x'}{x}): \sY_{1,q^{-1} x}  S_{1,x'}:
  \end{aligned}
  \label{eq:YS-commun}
\end{equation}
The total sum is $q_2$-difference which cancels after summation over $q_2^{\BZ}$ shifts entering definition of screening charge (\ref{eq:St}).
 This is the consequence of the Ward identity in Sec.~\ref{eq:Ward}. 

\subsection{W-algebra of $A_1$-quiver}
Consequently the operator $\tT_{1,x}$ can be moved in the position in the radial-ordered operator-state presentation of the extended gauge theory partition state (\ref{eq:ZTSS})
\begin{equation}
\tT_{1,x} |1 \rangle =  \tT_{1,x} \sS_{1,x'} \sS_{1,x'{}'} \dots |1\rangle
 =  \sS_{1,x'} \tT_{1,x} \sS_{1, x'{}'} \dots |1\rangle =   \sS_{1,x'}
 \sS_{1,x'{}' } \tT_{1,x} \dots |1\rangle
\end{equation} 
 The operators $\sS_{i,x'}$ can be thought as exponentiated Hamiltonians of the $q_1,q_2$-deformed CFT.
 The commutant of the Hamiltonians is the conserved current $\tT_{1,x}$ which is regular
\begin{equation}
  \partial_{\bar x} \tT_{1,x} = 0
\end{equation}
Consequently, $T_{1,x}$ has well defined, time-radial independent, modes
\begin{equation}
  T_{1,x} = \sum_{p \in \BZ} T_{1,[p]} x^{-p} 
\end{equation}

We define the algebra $W_{q_1,q_2}(A_1)$, a.k.a. the $q$-deformed Virasoro algebra, to be the subalgebra in $\bH$ generated by the modes of the conserved current $T_{1,x}$.
 This definition is in the exact agreement with Shiraishi et al.~\cite{Shiraishi:1995rp} and Frenkel--Reshetikhin \cite{Frenkel:1997}.

\subsection{W-algebra of quiver: definition}
The definition of the state $|Z_{\sT} \rangle$  (\ref{eq:ZTSS}) implies that the current $T_{i,x}$ is regular 
\begin{equation}
\label{eq:Tregg}
  \p_{\bar x} T_{i,x} |Z_{\sT} \rangle = 0
\end{equation}
if it commutes with all screening operators:
\begin{equation}
  [T_{i,x}, \sS_{j, x'}] = 0 \qquad j \in \Gamma_0, \qquad x' \in \CalX_{j}
\end{equation}
This explains isomorphism between the gauge theoretic construction of $q_1q_2$-characters \cite{Nekrasov:2013xda,Nekrasov:2015wsu} and definition of $W_{q_1,q_2}$-algebras \cite{Shiraishi:1995rp, Frenkel:1997,Frenkel:2010wd} as the algebra generated by currents $(T_{i,x})_{i \in \Gamma_0}$ which are defined as commutants of screening charges $(\sS_i)_{i \in \Gamma_0}$ in the vertex operator algebra defined by the free fields from Heisenberg algebra  $\bH$  and expressed as 
 \begin{equation}
   T_{i,x} = \sY_{i,x} + \dots 
 \end{equation}

 We define in the same way the W-algebra $W(\Gamma, \spacetime)$ for generic quiver $\Gamma$ with generalized even symmetric Borcherds--Kac--Moody--Cartan matrix, mass deformed by $\mu: \Gamma_1 \to \BC^{\times}$, as in equation (\ref{eq:massC}), and for $\spacetime = \BC_{q_1, q_2}$ as the algebra generated by currents $T_{i,x}$ commuting with all screening charges $(\sS_{i,x})_{i \in \Gamma_0}$, or equivalently, regular on the higher times extended gauge theory state (\ref{eq:Tregg}).
 We expect to generalize the definition for more general and possibly higher dimensional varieties $\spacetime$. 

\section{Examples}

We consider a few examples to illustrate the equivalence between gauge-theory formalism~\cite{Nekrasov:2012xe,Nekrasov:2013xda,Nekrasov:2015wsu} and the operator formalism~\cite{Shiraishi:1995rp,Frenkel:1997,Frenkel:1998,Frenkel:2010wd, Frenkel:2013uda}.

\subsection{Commutator of $T$ and $\sS$ vanishing: general quiver, local reflection}

 Suppose that there is no edge loop from a node $i$ to itself and consider 
\begin{equation}
T_{i,x} =  \sY_{i,x} \, + :\sY_{i, q^{-1} x}^{-1} \prod_{e: i \to j}
  \sY_{j, \mu_e^{-1} x }
    \prod_{e: j \to i } \sY_{j, q^{-1} \mu_{e} x}: + \dots 
    \label{eq:local_cancellation}
\end{equation}
The vanishing of commutator
\begin{equation}
\label{eq:TS}
  [T_{i,x}, \sS_{i,x'}]  = 0
\end{equation}
follows from (\ref{eq:YScommutator}) and the relation
\begin{equation}
 q_1
  :
  \sY_{i, q^{-1} x}^{-1}
  \left(
  \prod_{e: i \to j} \sY_{j, \mu_e^{-1} x }
  \prod_{e: j \to i } \sY_{j, q^{-1} \mu_{e} x}
  \right)
  S_{i,q^{-1}x}
  :
  \, = \, : \sY_{i,x} S_{i,q_1^{-1}x}:
\end{equation}
Indeed, this relation is equivalent to  
\begin{equation}
\label{eq:cc}
q_1^{-1}
: \sY_{i,x}  \sY_{i, q^{-1} x}   \left(
  \prod_{e: i \to j} \sY_{j, \mu_e^{-1} x }
  \prod_{e: j \to i } \sY_{j, q^{-1} \mu_{e} x}
  \right)^{-1} : \, = \, : S_{i, q^{-1} x} S_{i, q_1^{-1} x}^{-1} : 
\end{equation}
which simply expresses the defining relation (\ref{eq:yins}) between $\sY_{i,x}$ and $S_{j,x}$ in the
exponentiated form: the field $y_i(x)$ (of the Cartan weight type) is the $q_2$-derivative of the
field $s_i(x)$ (of the Cartan root type). Namely, the relation (\ref{eq:cc})
is the identity 
\begin{equation}
 \log q_2^{-1} s_{k,0} \tilde c_{kj}^{[0]} c_{ji}^{[0]}  
  + \sum_{p \neq 0} (1 - q_2^{-p}) q^{p} s_{k,p} \tilde c_{kj}^{[p]} c_{ji}^{[p]} x^{-p}
  = (\log(xq^{-1}) - \log(x q_1^{-1})) s_{i,0}
  + \sum_{p \neq 0} (q^{p} -q_1^{p}) s_{i,p} x^{-p}  
\end{equation}
thanks to the definition of the $\mu$-dependent Cartan matrix $c_{ij}$ (\ref{eq:massC})
and its inverse $\tilde c_{ij}$ so that $\tilde c_{kj}^{[p]} c_{ji}^{[p]}  = \delta_{ki}$. 
This leads to
\begin{equation}
  \left[
 : \sY_{i,q^{-1}x}^{-1}
  \prod_{e:i \to j} \sY_{j,\mu_e^{-1}x}
  \prod_{e:j \to i} \sY_{j,q^{-1}\mu_e x}: \, ,
  S_{i,x'}
  \right]
  = (q_1^{-1} -1) \delta (q \frac{x'}{x}): \sY_{1,q^{-1} x}  S_{1,x'}:  
\end{equation}
and therefore (\ref{eq:TS}) holds at the level of the first two terms. 
 The second term contains the $\sY_{j}$ fields for the nodes $j$ linked to the node $i$.
 This term might give potential singularities, or, equivalently, $\delta$-functions in the commutators associated to the $\sS_{i}$ operators. Then one needs to continue to apply Weyl reflections to generate terms which cancel the singularities. 
The algebraic structure is associated to highest weight Verma module of the generalized 
Borcherds--Kac--Moody algebra $\g_{\Gamma}$. 
If $\Gamma$ is of finite Dynkin type the process terminates, the associated Verma module is finite-dimensional.
 The finite-dimensional case was studied in details in \cite{Frenkel:2010wd}. 

More generally, for infinite-dimensional Verma module,  the recursive
algorithm is also applicable which builds a tree starting from the root
node $\sY_{i,x}$.  The vertices of the tree are monomials in the
$T_{i,x}$ current, and the edges are colored by the nodes $i$ of the
quiver. Two monomials are linked by edge of color $i$ if they are
related by the local reflection move (\ref{eq:local_cancellation}).
The algorithm can be computerized. 

Alternatively, Nekrasov presented closed formula \cite{Nekrasov:2015wsu}
which express the $q_1q_2$-characters in terms of geometry of Nakajima's
quiver variety \cite{Nakajima1994,Nakajima:1999}. This formula can be
thought as $q_2$-deformation of original Nakajima's construction of
$q$-characters of  $\mathbf{U}_{q}(L \mathfrak{g}_{\Gamma})$ from the
$q$-equivariant K-theory on the quiver variety
$\frakM_{\bw,\bv}^{\mathrm{Nak}}=T^{*}_{q}\frakM_{\bw,\bv}$
\cite{Nakajima:2004}.  The formula
in \cite{Nekrasov:2015wsu} amounts to replacing Euler
characteristic of $T^{*}_{q}\frakM_{\bw,\bv}$  by the $q_2$-equivariant Euler class of the tangent
bundle to $T^{*}_{q}\frakM_{\bw,\bv}$ so effectively to the integration
over $\Pi T_{q_2} T^{*}_{q}\frakM_{\bw,\bv}$. 
 Here $\bw: \Gamma_0 \to \BZ$ labels the components of the highest weight
in the basis of fundamental weights, and $\bv: \Gamma_0 \to \BZ$ labels
the components of a positive root in the basis of simple roots which is
added to the highest weight to get a weight at the level $\sum_{i \in
  \Gamma_0} {\bv_i}$ in the Verma module. 

\subsection{Higher weight currents}

 Conjecturally, quiver W-algebra is completely generated by the
fundamental currents 
\begin{equation}
  T_{i,x} = \sY_{i,x} + \dots 
\end{equation}
 However, higher weight currents $ T^{\bw}_{w,x}$ can be defined where to each node $i$ we assign vector space $\bW$ of dimension $\bw$ and the character 
\begin{equation}
  W_{i} = \sum_{\omega=1}^{\bw_i}  w_{i,\omega} \qquad w_{i, \omega} \in \BC^{\times}
\end{equation}
with the first term 
\begin{equation}
 T^{\bw}_{w,x} = \,
  : \prod_{i \in \Gamma_0} \prod_{\omega=1}^{\bw_i}
  \sY_{i, w_{i,\omega} x} : + \dots 
\end{equation}

In the finite-dimensional and  irreducible modules of higher weights can
be found in the tensor product of the $i$-fundamental modules with
weights $ \bw_{i} = 1, \bw_{j \neq i} = 0$. In the not $q$-deformed case,
usually the tensor product of fundamental modules decomposes into
several irreducible components. For example, for $\mathfrak{sl}_2$ we
have $\BC^2 \otimes \BC^2 = \BC^3 \oplus \BC^1$. 
This does not hold after $q$-deformation.  For generic weights $w$ the tensor product is irreducible.

\subsection{Higher weight current in the $A_1$ example}
In the example of $A_1$ quiver the higher weight  current $T^{\bw}_{w, x}$ with
$\bw_1 \in \BZ_{>0}$ for generic weights $(w_{1,1}, \dots, w_{1,\bw_1})$ contains
$2^{\bw_1}$ terms \cite{Nekrasov:2015wsu} coming from the cohomologies of
Nakajima's quiver variety which in this case are $\amalg_{\bv \leq \bw} T^{*}\mathbf{Gr}(\bw, \bv)$.
This higher weight character current $T^{\bw}_{w, x}$ is elementary to compute from
the  free-field formalism and normal ordering given the fundamental
current $T_{1,x}$ in equation (\ref{eq:A1-current})

 Consider the product
\begin{equation}
  T_{1,w_1 x} T_{1, w_2 x}  = (\sY_{1,xw_1} + \sY^{-1}_{1,
    q^{-1} w_1 x}) (\sY_{1,xw_2} + \sY^{-1}_{1,
    q^{-1} w_2 x})
\end{equation}
The normal ordering is computed using the commutator from (\ref{eq:trel})
\begin{equation}
  [y_{i,p}, y_{j,-p}] \stackrel{p>0}{=} -\frac{1}{p}
  (1-q_1^{p})(1-q_2^{p}) \tilde c^{[-p]}_{ji}
\end{equation}
with the result 
\begin{multline}
    T_{1,w_1 x} T_{1, w_2 x} = f(w_2/w_1)^{-1}\Big(
 : \sY_{1,w_1 x} \sY_{1,w_2 x}: \\+
 \msS(w_1/w_2) : \sY_{1, w_1 x}
 \sY_{1,q^{-1} w_2 x}^{-1}: + \msS(w_2/w_1) :\sY_{1,q^{-1} w_1 x}^{-1}  \sY_{1, w_2 x}:   \\ +
 :\sY_{1, q^{-1} w_1 x}^{-1}   \sY_{1,q^{-1} w_2 x}^{-1}: \Big)
\end{multline}
where the scalar prefactor 
\begin{equation}
\label{eq:ffun}  f(w) =\exp\left( \sum_{p=1}^\infty \frac{1}{p} \frac{ (1-q_1^{p})(1 - q_2^{p})}{ 1 +
      q^{p}} w^{p} \right)
\end{equation}
is in agreement with the function $f(w)$ generating the commutation
relations for $W_{q_1, q_2}(A_1)$ current $T_1(x)$  in Shiraishi et al.~\cite{Shiraishi:1995rp} 
and the permutation factor $\msS(u)$  is in agreement with formulae for higher 
$qq$-characters in \cite{Nekrasov:2015wsu}
\begin{equation}
\label{eq:sm}
 \msS(w) = \frac{(1 - q_1 w)(1 - q_2 w)}{(1 - q w)(1 - w)}
\end{equation}
which comes from the equivariant Euler characteristic (or its K-theory
version) of the fixed point in  $\Pi T_{q_2} T^{*}_{q_1} \BP^1$ where $T^{*}_{q_1} \BP^1
= \frakM^{\mathrm{Nak}}_{\bw=2, \bv=1}$.
This relation leads to
\begin{align}
 f\left( \frac{w_2}{w_1} \right) T_{1,w_1 x} T_{1,w_2 x}
 - f\left( \frac{w_1}{w_2} \right) T_{1,w_1 x} T_{1,w_2 x}
 & = \frac{(1-q_1)(1-q_2)}{1-q}
 \left(
 \delta\left( q \frac{w_1}{w_2} \right)
 - \delta\left( q \frac{w_2}{w_1} \right) 
 \right)
 \, ,
\end{align}
which determines the algebraic relation for the modes $(T_{1,[p]})_{p \in \BZ}$.
We remark $f(w) f(qw) = \msS(w)$.

The degree $\sw$ current is similarly computed 
\begin{eqnarray}
 T_{1,x}^{[\sw]}
  & = &
  :\sY_{1,w_1 x} \sY_{1,w_2 x} \cdots \sY_{1,w_\sw x}: + \cdots
  \nonumber \\
 & = &
  \sum_{I \cup J = \{1 \ldots \sw\}}
  \prod_{i \in I, j \in J} 
  \msS\left( \frac{w_i}{w_j} \right)
  :
  \prod_{i \in I} \sY_{1,w_i x}
  \prod_{j \in J} \sY^{-1}_{1,q^{-1} w_j x}
  :
\end{eqnarray}
in agreement with \cite{Nekrasov:2015wsu}.
The $\msS$ factor becomes trivial in the limit $q_{2} \to 1$ and the ordinary formulae for $q$-character is recovered~\cite{Frenkel:1998, Frenkel:2013uda, Nekrasov:2013xda}.

\subsection{Degeneration and derivatives}
 
 By definition, vertex operator algebra involves expressions in fields and their derivatives.
 Hence we shall expect appearance of the derivatives when two vertex operators fuse. 

So consider slightly more general situation of W-algebra currents with local structure
\begin{eqnarray}
 :\sY_{i,x} \sY_{i,ux}:
  & + & \msS(u)
  :\frac{\sY_{i,ux}}{\sY_{i,q^{-1} x}}
  \left(
  \prod_{e:i \to j} \sY_{j,\mu_{e}^{-1} x}
  \prod_{e:j \to i} \sY_{j,\mu_e q^{-1} x}
  \right)
  :
  \nonumber \\
 & + &
  \msS(u^{-1})
  :\frac{\sY_{i,x}}{\sY_{i,u q^{-1} x}}
  \left(
  \prod_{e:i \to j} \sY_{j,u\mu_{e}^{-1} x}
  \prod_{e:j \to i} \sY_{j,u\mu_e q^{-1} x}
  \right)
  :
  \nonumber \\
 & + &
  :
  \sY^{-1}_{i,q^{-1} x} \sY^{-1}_{i,u q^{-1} x}
  \left(
  \prod_{e:i \to j} \sY_{j,\mu_{e}^{-1} x} \sY_{j,u\mu_{e}^{-1} x}
  \prod_{e:j \to i} \sY_{j,\mu_e q^{-1} x} \sY_{j,u\mu_e q^{-1} x}
  \right)  
  :
\end{eqnarray}
Taking the collision limit $u \to 1$, this yields a derivative term
\begin{align}
 \sY^2_{i,x} \,
 & + 
 :
 \frac{\sY_{i,x}}{\sY_{i,q^{-1} x}}
 \prod_{e:i \to j} \sY_{j,\mu_e^{-1} x}
 \prod_{e:j \to i} \sY_{j,\mu_e q^{-1} x}
 \nonumber \\
 & \quad \times
  \left(
   \fc(q_1,q_2) - \frac{(1-q_1)(1-q_2)}{1-q}
 \frac{\partial}{\partial \log x}
 \log \left(
 \frac{\sY_{i,x} \sY_{i,q^{-1} x}}
      {\prod_{e:i \to j} \sY_{j,\mu_e^{-1} x}
       \prod_{e:j \to i} \sY_{j,\mu_e q^{-1} x}}
 \right)
  \right) 
  :
  \nonumber \\
 & + 
  :
  \sY^{-2}_{i,q^{-1} x}
  \left(
  \prod_{e:i \to j} \sY_{j,\mu_e^{-1} x}
  \prod_{e:j \to i} \sY_{j,\mu_e q^{-1} x}
  \right)^2
  :
\end{align}
where the coefficient $\fc(q_1,q_2)$ is determined by
\begin{eqnarray}
 \fc(q_1,q_2) & = &
 \lim_{u \to 1}
 \left(
  \msS(u) + \msS(u^{-1})
 \right)
 \nonumber \\
 & = &
 \frac{1 - 6q_1q_2 + q_1^2q_2^2 + (1 + q_1q_2)(q_1 + q_2)}{(1 - q_1q_2)^2}
 \ \stackrel{q_{1,2} \to 1}{\longrightarrow} \ 2
 \, .
\end{eqnarray}
We can consider more higher collision, which involves correspondingly higher derivatives.

\subsection{Edge loop: $\widehat{A}_0$-example}

 Consider an example of a single node with a loop edge.
 This corresponds to $\CalN=2^*$ theory in 4d. 
 Let $\sn \in \BZ_{\ge 1}$ be the gauge group rank and $\mu \in \BC^\times$ be the (multiplicative; exponentiated) adjoint mass.
 The Cartan matrix is $(0)$ and the mass deformed Cartan matrix is 
\begin{equation}
 c = 1 + q^{-1} - \mu^{-1} - q^{-1} \mu = (1 - \mu^{-1})(1 - q^{-1} \mu)
\end{equation}
 The quantum affinization of the respective algebra by Nakajima's quiver construction is $\mathbf{U}_{q, \mu}(L \widehat{\mathfrak{gl}}_{1})$~\cite{Schiffmann:2012b,Schiffmann:2009,Varagnolo:1999} with $q$-character given by the sum over all partitions~\cite{Nekrasov:2013xda}.
 Here we consider $q_2$-deformation to recover W-algebra of $\widehat A_0$-quiver.

We need the commutation relation for the oscillator \eqref{eq:trel} 
\begin{equation}
 [y_{1,p},y_{1,p'}]
  =
  - \delta_{p+p',0} \frac{1}{p}
  \frac{(1-q_1^p)(1-q_2^p)}{(1-\mu^p)(1-q^p \mu^{-p})}
\end{equation}
Using this oscillator, we construct $W_{q_1,q_2}(\g_\Gamma)$ algebra associated with the affine quiver $\Gamma=\widehat{A}_0$.
In this case, the local pole cancellation structure is
\begin{eqnarray}
 & &
  \sY_{1,x}
  +
  \msS(\mu^{-1})
  :\sY_{1,q^{-1}x}^{-1} \sY_{1,\mu^{-1} x} \sY_{1,\mu q^{-1}x}:
  \label{eq:pole_cancellation_A0}
\end{eqnarray}
and the holomorphic current can be characterized by a single partition~\cite{Nekrasov:2013xda}
\begin{eqnarray}
 T_{1,x} & = &
  \sY_{1,x} 
    +
    \msS(\mu^{-1})
  :\sY_{1,q^{-1}x}^{-1} \sY_{1,\mu^{-1} x} \sY_{1,\mu q^{-1}x}:
  + \cdots
  \nonumber \\
 & = &
  \sum_{\lambda} \tilde{Z}_\lambda
  :\prod_{s\in\partial_+\lambda} \sY_{1,qx/\tilde{x}(s)}
   \prod_{s\in\partial_-\lambda} \sY^{-1}_{1,x/\tilde{x}(s)} :
\end{eqnarray}
where $\partial_+ \lambda$ and $\partial_- \lambda$ are the outer and inner boundary of the partition $\lambda$, and we define
\begin{equation}
 \tilde{x}(s) = (\mu^{-1}q)^{s_1-1} \mu^{s_2-1} q
\end{equation}
The combinatorial weight $\tilde{Z}_\lambda$ obeys
\begin{equation}
 \frac{\tilde{Z}_{\lambda'}}{\tilde{Z}_\lambda}
  =
  - \frac{(1-\mu^{-1}q_1)(1-\mu^{-1}q_2)}{(1-q_1)(1-q_2^{-1})}
  \frac{\tilde{\sY}_{q_1x}\tilde{\sY}_{q_2x}}
       {\tilde{\sY}'_{qx}\tilde{\sY}_{x}}
       \Bigg|_{x=\tilde{x}_k}
\end{equation}
where $\lambda'$ is the shifted partition $\lambda_k \to \lambda_k + 1$, and we define the ``dual'' function $\tilde{\sY}_x$ 
\begin{equation}
 \tilde{\sY}_x
  =
  \prod_{k=1}
  \frac{1-\tilde{x}_k/x}{1-\tilde{q}_1\tilde{x}_k/x}
\end{equation}
with the ``dual'' parameters
\begin{equation}
  \tilde{q}_1 = \mu^{-1} q
  \, , \qquad
  \tilde{q}_2 = \mu
  \, , \qquad
  \tilde{\mu} = q_2
  \, , \qquad
  \tilde{x}_k = \tilde{x}(k,\lambda_k+1)
\end{equation}
Here $\tilde{\sY}'$ is evaluated with the shifted configuration $\lambda'$.
Although this dual function also has poles, such a singularity is cancelled in the following combination
\begin{equation}
 \tilde{\sY}_x
  +
  \frac{(1 - \tilde\mu^{-1}\tilde{q}_1)(1 - \tilde\mu^{-1}\tilde{q}_2)}
       {(1 - \tilde\mu^{-1} \tilde{q})(1 - \tilde\mu^{-1})}
  \tilde{\sY}^{-1}_{\tilde{q}^{-1}x}
  \tilde{\sY}_{\tilde{\mu}^{-1} x} \tilde{\sY}_{\tilde{\mu}\tilde{q} x}
\end{equation}
This expression is equivalent to the original one
\eqref{eq:pole_cancellation_A0} in particular for the rank one theory.
Again, operator formalism of W-algebra is equivalent to
$q_2$-deformation of Nakajima's construction \cite{Nekrasov:2015wsu}. 

\subsection{W-algebra of hyperbolic quiver example}

We consider examples of the hyperbolic quiver, where the determinant of the corresponding Cartan matrix is negative.
The simplest example is the quiver having a single node with two loop edges:
\begin{equation}
    \begin{tikzcd}
     \bullet \arrow[loop left]{r} \arrow[loop right]{r}
    \end{tikzcd} \qquad c = - (2)
\end{equation}
Let $\mu_{1,2} \in \BC^\times$ be the mass parameter associated with the edges, and the mass deformed Cartan matrix is given by
\begin{equation}
 c = 1 + q^{-1} - \mu_1^{-1} - \mu_1 q^{-1} - \mu_2^{-1} - \mu_2 q^{-1} 
\end{equation}
Since the local pole cancellation occurs in the following combination
\begin{equation}
 \sY_{1,x} + \,
  \msS(\mu_1^{-1}) \msS(\mu_2^{-1})
  :\sY_{1,q^{-1}x}^{-1}
  \sY_{1,\mu_1^{-1}x} \sY_{1,\mu_1 q^{-1}x}
  \sY_{1,\mu_2^{-1}x} \sY_{1,\mu_2 q^{-1}x}:
\end{equation}
the first few terms of the holomorphic current are given by
\begin{eqnarray}
 T_{1,x} & = &
  \sY_{1,x} + \msS(\mu_1^{-1}) \msS(\mu_2^{-1})
  :\sY_{1,q^{-1}x}^{-1}
  \sY_{1,\mu_1^{-1}x} \sY_{1,\mu_1 q^{-1}x}
  \sY_{1,\mu_2^{-1}x} \sY_{1,\mu_2 q^{-1}x}:
  \nonumber \\
 & & +
  \left( \msS(\mu_1^{-1}) \msS(\mu_2^{-1}) \right)^2
  \Bigg[
  \nonumber \\
 & & \qquad
  \msS(\mu_1^2 q^{-1}) \msS(\mu_1\mu_2^{-1}) \msS(\mu_1 \mu_2 q^{-1})
  :\frac{\sY_{1,\mu_1^{-2}x} \sY_{1,\mu_1 q^{-1}x}
   \sY_{1,\mu_1^{-1} \mu_2^{-1} x} \sY_{1,\mu_1^{-1} \mu_2 q^{-1} x}
   \sY_{1,\mu_2^{-1} x} \sY_{1,\mu_2 q^{-1} x}}
       {\sY_{1,\mu_1^{-1} q^{-1} x}}:
       \nonumber \\
 & & \qquad
  + \, \msS(\mu_1^{-2} q) \msS(\mu_1^{-1}\mu_2^{-1}q) \msS(\mu_1^{-1}\mu_2)
  :\frac{\sY_{1,\mu_1^{-1}x} \sY_{1,\mu_1^2 q^{-2}x}
   \sY_{1,\mu_1 \mu_2^{-1} x} \sY_{1,\mu_1 \mu_2 q^{-1} x}
   \sY_{1,\mu_2^{-1} x} \sY_{1,\mu_2 q^{-1} x}}
   {\sY_{1,\mu_1 q^{-2} x}}:
   \nonumber \\
 & & \qquad
  + \, ( 1 \leftrightarrow 2 )
  \Bigg]
  + \cdots
\end{eqnarray}
We can see a cancellation of factors, which is similar to $\hat{A}_0$ theory, and thus there is no colliding term, e.g. $\sY_{1,*}^2$, in a numerator.

Next example is a rank two quiver with three arrows:
\begin{equation}
    \begin{tikzcd}
      \bullet \arrow[bend left]{r} \arrow{r}  \arrow[bend right]{r} & \bullet
    \end{tikzcd} \qquad c =
    \begin{pmatrix}
      2 & -3\\
    -3 & 2 
    \end{pmatrix}
\end{equation}
Let us assign three mass parameters $\mu_{1,2,3}$ to the arrows, and then the local cancellation is
\begin{equation}
 \sY_{1,x} + 
  :\frac{\sY_{2,\mu_1^{-1}x} \sY_{2,\mu_2^{-1} x} \sY_{2,\mu_3^{-1} x}}
        {\sY_{1,q^{-1} x}}:
       \qquad
 \sY_{2,x} + 
 :\frac{\sY_{1,\mu_1 q^{-1}x} \sY_{1,\mu_2 q^{-1} x} \sY_{1,\mu_3 q^{-1} x}}
       {\sY_{2,q^{-1} x}} :
\end{equation}
The holomorphic current becomes
\begin{eqnarray}
 T_{1,x}
  & = &
  \sY_{1,x} + 
  :\frac{\sY_{2,\mu_1^{-1}x} \sY_{2,\mu_2^{-1} x} \sY_{2,\mu_3^{-1} x}}
        {\sY_{1,q^{-1} x}}:
  \nonumber \\
 && + \Bigg[
  \msS(\mu_1 \mu_2^{-2}) \msS(\mu_1 \mu_3^{-1})
  :\frac{\sY_{1,\mu_1^{-1}\mu_2 q^{-1}} \sY_{1,\mu_1^{-1}\mu_3 q^{-1}}
         \sY_{2,\mu_2^{-1} x} \sY_{2,\mu_3^{-1} x}}
	{\sY_{2,\mu_1^{-1} q^{-1}x}}
  :
  \nonumber \\
 && \qquad +
  \msS(\mu_2 \mu_1^{-2}) \msS(\mu_2 \mu_3^{-1})
  :\frac{\sY_{1,\mu_2^{-1}\mu_1 q^{-1}} \sY_{1,\mu_2^{-1}\mu_3 q^{-1}}
         \sY_{2,\mu_1^{-1} x} \sY_{2,\mu_3^{-1} x}}
	{\sY_{2,\mu_2^{-1} q^{-1}x}}
  :  
  \nonumber \\
 && \qquad +
  \msS(\mu_3 \mu_1^{-2}) \msS(\mu_3 \mu_1^{-1})
  :\frac{\sY_{1,\mu_3^{-1}\mu_1 q^{-1}} \sY_{1,\mu_3^{-1}\mu_2 q^{-1}}
         \sY_{2,\mu_1^{-1} x} \sY_{2,\mu_2^{-1} x}}
	{\sY_{2,\mu_3^{-1} q^{-1}x}}
  :  
  \Bigg]
  \nonumber \\
 && \qquad + \cdots  
\end{eqnarray}
The other current $T_{2,x}$ is obtained in the same way.
Similarly it is expected that there is no collision term in these
holomorphic currents.

\section{Applications}
\subsection{Toda scaling limit}

 In the scaling limit $q_1 \to 1, q_2 \to 1, \fq_i \to 1$, and $\log_{q_2} q_1, \log_{q_2} \fq_i$ are finite, the free field commutation relations (\ref{eq:scomp})(\ref{eq:scom}) turn into
\begin{equation}
\label{eq:scomp1}
\begin{aligned}
{} [ s_{i,p}, s_{j,p'}] &=- \delta_{p+p',0}\frac{1}{p} \beta c_{ji}^{[0]}, \qquad p > 0 \\
[ \tilde s_{i,0} , s_{j,p}] &= -\beta  \delta_{0,p} c_{ji}^{[0]}
\end{aligned}
\end{equation}
This limit was studied in details in section 4.1 of~\cite{Frenkel:1997}. 
In terms of the parameter $b^2 = - \beta$ the vertex operator (\ref{eq:S_i}) can be written as 
\begin{equation}
 S_i(x) = :e^{b \phi_i(x)} :
\end{equation}
where $\phi_i(x)$ is the free boson that takes value in the Cartan part of $\g_{\Gamma}$ with canonical commutation relations defined by the bilinear form with matrix $(c_{ij})$ in the basis of simple roots.
 Hence $S_i(x)$ are vertex primary operators of \emph{Kac--Moody $\g_{\Gamma}$-Toda field theory} on punctured disc $\BC^{\times}_{x}$.
In the same scaling limit we find from (\ref{eq:yins}) that the field~$y_i(x)=-\ep_2 b \p \phi_j(x) \tilde c_{ji}$ is also primary. 

For example,  in the $\ep_2$-expansion of $T_1(x)$ for $A_1$-quiver 
\begin{equation}
 T_1(x)
  = \, :e^{y(x)}: + :e^{-y(xq^{-1})}: \,
  = 2 + \frac{1}{4} \ep^2 b^2( ( \p \phi )^2  - (b + b^{-1})\p^2 \phi) + \dots 
\end{equation} 
we find the stress-energy Virasoro current of the free field $\phi$ with background charge and the central charge 
\begin{equation}
  c = 1 + 6 (b + b^{-1})^2
\end{equation}

\subsection{Affine type}

If $\g_{\Gamma}$ is of affine type, the $\g_{\Gamma}$-Toda is affine Toda.
 For example, the scaling limit of the W-algebra defined by the quiver
\begin{equation}
    \begin{tikzcd}
      \bullet \arrow[bend left]{r}   \arrow[bend right]{r} & \bullet
    \end{tikzcd} \qquad c =
    \begin{pmatrix}
      2 &-2\\
    -2 & 2 
    \end{pmatrix}
\end{equation}
with $\g_{\Gamma} = A_1^{(1)}$ describes quantum sin(h)-Gordon theory on punctured disc $\BC^{\times}_{x}$.

\subsection{Nahm transform}
 The $\g_{\Gamma}$-Toda theory specializes to the finite Toda if $\g_{\Gamma}$ is of finite type.
 For $\mathfrak{sl}_{r}$-quiver with $n$ colors at each node the $\mathfrak{sl}_{r}$-Toda is Nahm dual to the $\mathfrak{sl}_n$-Toda proposed in \cite{Alday:2009aq,Wyllard:2009hg}.

\bibliographystyle{utphysurl} \bibliography{wquiver} 
\end{document}